\documentclass[%
 reprint,
 showkeys,
 onecolumn,
 amsmath,amssymb,
 aps,
prb,
floatfix,
]{revtex4-2}

\usepackage{graphicx}
\usepackage{dcolumn}
\usepackage[version=4]{mhchem}
\usepackage{tabularray}
\usepackage{xcolor}
\usepackage{colortbl}
\usepackage{breakcites}
\usepackage{url}
\usepackage{xurl}
\usepackage{soul}
\usepackage{ragged2e}

\usepackage{array,makecell}
\newcolumntype{F}[1]{%
	>{\raggedright\arraybackslash\hspace{0pt}}p{#1}}%
\newcolumntype{T}[1]{%
	>{\centering\arraybackslash\hspace{0pt}}p{#1}}%
	
\usepackage{ninecolors}  

\usepackage[slovak,english]{babel}

\definecolor{pink}{rgb}{1,0,1}
\definecolor{ruzz}{rgb}{1,0,0.5}
\definecolor{ruz}{rgb}{1,0,0.5}
\definecolor{tmod}{RGB}{0,50,150}
\definecolor{tzel}{RGB}{0,200,200}
\definecolor{zel}{RGB}{0,100,50}
\definecolor{fialova}{rgb}{0.5,0,0.5}
\definecolor{white}{rgb}{1.0,1.0,1.0}
\definecolor{tblue}{RGB}{0,0,130}
\definecolor{seda}{RGB}{235,235,235}
\definecolor{sseda}{RGB}{248,248,248}
\definecolor{azur}{RGB}{212,255,230}
\definecolor{grey}{RGB}{255,240,240}
\definecolor{ye}{RGB}{255,255,0}
\colorlet{lightye}{ye!37}
\definecolor{lavender}{rgb}{0.9, 0.9, 0.98}

\setcitestyle{citesep={,}}
\usepackage{hyperref}
\hypersetup{
	colorlinks = true,
	linkcolor = ruzz,
	citecolor = blue,
	linkbordercolor = red,
	citebordercolor = green,
	urlcolor= fialova
}

\newcommand{\hlc}[2][yellow]{{\sethlcolor{#1}\hl{#2}}}

\begin{document}

\preprint{APS/123-QED}

\title{Family of $\mathrm{\mathbf{(N \times H)}}$-Polytypes with \ce{La2WO6}-Related Stoichiometry}

\author{Eva Posp\' i\v silov\' a}
\email{pospiloveva255@gmail.com}
\email{e.pospisilova@savba.sk}
\affiliation{Institute of Materials and Machine Mechanics, Slovak Academy of Sciences, Bratislava, 84513, D\' ubravsk\' a~cesta~9}
\affiliation{Institute of Physics, Slovak Academy of Sciences, Bratislava, 84511, D\' ubravsk\' a~cesta~11, Slovakia}

\author{Marek Mihalkovi\v{c}}
\email{marek.mihalkovic@savba.sk}
\affiliation{Institute of Physics, Slovak Academy of Sciences, Bratislava, 84511, D\' ubravsk\' a~cesta~11, Slovakia}

\author{Na\v da Beronsk\'{a}}
\email{nada.beronska@savba.sk}
\affiliation{Institute of Materials and Machine Mechanics, Slovak Academy of Sciences, Bratislava, 84513, D\' ubravsk\' a~cesta~9}

\author{Marek Gebura}
\email{marek.gebura@savba.sk}
\affiliation{Institute of Materials and Machine Mechanics, Slovak Academy of Sciences, Bratislava, 84513, D\' ubravsk\' a~cesta~9}

\date{\today}

\begin{abstract}
Structure and chemical composition of non-stoichiometric $\mathrm{(N \times H)}$-polytypes, $\mathrm{N} \in \{3,\,4,\,5,\,6,\,7\}$, belonging to the family of~\ce{La2WO6}-related tungstates, are presented and the basic rules behind their construction are formulated. The polytypes are validated from the point of view of the internal energy per~atom against all the competing La-O-W compounds by means of the ternary convex hull computed by~DFT. According to DFT, the ground state of the \ce{La2WO6} family surprisingly turned out to~be the as-yet \emph{hypothetical} tungstate \ce{La2WO6}.oP36, isostructural with the known compound \ce{Gd2WO6}, closely related to the basic building H-block of the polytypes. One more candidate for an experimentally accessible low-energy structure, \ce{La2WO6}.mC72, derived from the already observed \ce{Sm2MoO6}, has been identified to lie only $\Delta E \approx 5\:\mathrm{meV/at.} \cong 60\:\mathrm{K}$ above the convex envelope.
\end{abstract}


\maketitle



\section{\label{intro}Introduction}

\indent Lanthanum tungstates, i.e. ternary oxides with La-O-W composition, such as \ce{La6WO12} or \ce{La2W2O9}, are studied mainly for their application as electrolytes in proton conducting Solid Oxide Fuel Cells (SOFC), \cite{magraso2013,cao,kojo,la10w2o21}. At elevated temperatures they develop significant electronic conductivity, making them suitable for dense hydrogen gas separation membranes, \cite{magraso2012,magraso2013}. In addition, fast oxide-ion conductors from the LAMOX family (\ce{La10W2O21}, $\beta$-\ce{La2W2O9}) have found applications in oxygen sensors and oxygen pumping devices, \cite{229-nature,229-2000,la10w2o21}. La oxides and especially weak La bonds with W atoms on the welding electrode surface lower its work function significantly, thus reducing the electrode operating voltage, temperature, erosion rate and enhancing its longevity, \cite{ele1}. \\
\indent The most striking feature of the La-O-W phase diagram is that all ternary compounds fall into the line segment connecting a couple of binary oxides: \ce{WO3} and \ce{La2O3}, because they are experimentally prepared from~the two powders. The only degree of freedom is thus the exact molar ratio of the two constituents. This enables the classification of La tungstates into several families according to~the $\mathrm{La:W}$ ratio: \ce{La6WO12} ($\mathrm{La:W} = 6:1$), \ce{La6W2O15} ($\mathrm{La:W} = 3:1$), \ce{La2WO6} ($\mathrm{La:W} = 2:1$), \ce{La2W2O9} ($\mathrm{La:W} = 1:1$), \ce{La2W3O12} ($\mathrm{La:W} = 2:3$), each exhibiting a particular crystalline structure, subjected to subtle modulations under slight deviations of the stoichiometry (i.e. of the $\mathrm{La:W}$ ratio). The value of~the $\mathrm{La:W}$ ratio thus serves as an~identifier of a tungstate family, determines its crystal structure and all the interesting physical properties. \\
\indent In the present contribution, we would like to focus our attention on the family of \ce{La2WO6}-related tungstates, forming so-called $\mathrm{(N\times H)}$-\emph{polytypes}. The polytypes are formed by slight structural modulations of the basic hexagonal block, hereafter H-block. They have been the subject of the studies of M.-H. Chambrier et al., \cite{la18w10o57,a-la2wo6,b-la2wo6,MHCh,la10w2o21,chambrier-thesis} who mapped experimentally the whole La-O-W equilibrium phase diagram, and of the group of N. E. Novikova et al., \cite{5H}, following the preceding work of V. K. Yanovskii \& V. I. Voronkova, \cite{voronkova1,voronkova2}, M. M. Ivanova et al., \cite{ivanova}, and A. Magras\'{o} et al., \cite{la-o-w-equi}, and the authoritative first work on La-O-W diagram by M. Yoshimura and A. Rouanet,~\cite{yoshimura}. \\
\indent All the above-mentioned papers are purely experimental, based on ab initio structure determination from X-ray and neutron diffraction experiments. None of them provided an explanation of the structural modulation of the H-block leading to the $\mathrm{(N\times H)}$-polytypes. The studies of well-equilibrated La tungstates by the group of M.-H. Chambrier have never been backed up by any theoretical predictions, which might have saved a large volume of experimental material, time and resources. Besides economic and ecological advantages, density functional theory (DFT) energy minimization can be exceptionally precise in more advanced structure and composition refinement. DFT can surpass even state-of-the-art diffraction methods, which offer only an indirect way of measuring composition by complicated and unreliable multi-component fits, in contrast to direct -- e.g. energy-dispersive X-ray spectroscopy -- probes. \\ 
\indent There is a fundamental difficulty of averaging over disorder (AOD), which is clearly beyond the scope of any diffraction method. This is particularly relevant to the $\mathrm{(N\times H)}$-polytypes, which are always a mixture of at least two distinct members, namely the most frequently encountered 5H- and 6H-polytypes, since all attempts to synthesize them as single crystals were unsuccessful, \cite{la18w10o57,voronkova1,voronkova2,ce18w10o57}. Hence diffraction methods always yield W atomic positions averaged over several polytypes and consequently multiple W positions with rather low occupancy. Only \emph{ab initio} numerical approaches, such as DFT, are capable of deciphering occupancy \emph{correlations} that are lost in AOD. In~this sense, DFT and experimental diffraction methods complement each other. \\
\indent What is more, experimentalists do not have in general the information about internal-energy differences between~any given compounds at their disposal; except for designated calorimetric measurements, there is no guarantee that the compound they prepared is actually in true thermodynamic equilibrium. \\     
\indent In the field of computational chemistry the construction of the convex hull covering \emph{all} known stable compounds of the given elements is \emph{the only} way to adequately evaluate the stability of any compound with~respect to~its chemical decomposition into constituent species. Therefore, based on all the above-mentioned papers and on all the accessible relevant experimental phase diagrams, we would like to~present alternative and supplementary DFT results on La-O-W oxide stability. By means of~the reassessed equilibrium zero-temperature zero-pressure convex envelope we would like to~single out the most stable structures and stoichiometries and elucidate the structural modifications behind the $\mathrm{(N\times H)}$-polytypes. To the best of~our knowledge, the only theoretical evaluation of the La-O-W convex hull -- though without~any given structural information -- is available in~\cite{oqmd1,oqmd2}.


\section{\label{methods}Methods}

\subsection{\label{meth:basic}Basic Setup \& Convergence}

\indent In order to compute the La-O-W convex hull at $T = 0\:\mathrm{K}$, $p = 0\:\mathrm{kbar}$, \textit{ab initio} Density Functional Theory (DFT) setup was employed as implemented in the Plane-Wave package \texttt{VASP} (Vienna \emph{ab initio} Simulation Package), version 6.5.0, \cite{vvasp}, using Projector Augmented-Wave (PAW) method potentials, \cite{paw}, and the Perdew-Burke-Ernzerhof density functional ``PBE'', \cite{pbe}, utilizing the Generalized Gradient Approximation (GGA). O: $s^{2}p^{4}$ and W: $6s^{2}5d^{4}$ electrons were treated as valence. For La, Kr$4d$ were regarded as core electrons; $5s^{2}6s^{2}5p^{6}5d^{1}$ were treated as valence. As~a~default setup, we use \texttt{VASP} ``Accurate'' setting, and converge the energy with respect to the K-point mesh density to~$1\:\mathrm{meV/at.}$ accuracy. Additionally, all internal-energy results were converged with respect to the energy cutoff, namely \texttt{ENCUT} $ = 550\:\mathrm{eV}$ turned out to be sufficient. All sites in the unit cell along~with~the unit cell dimensions were relaxed using a conjugate gradient algorithm to minimize the energy with~an~atomic force tolerance of $0.005\:\mathrm{eV/}$\AA\, and a total energy precision of~$10^{-5}\:\mathrm{eV}$. \\
\indent Two setups were compared:

\vspace{-0.0cm}

\begin{enumerate}
\item \texttt{GGA=91}, the PW91 Perdew-Wang functional, \vspace{-0.0cm}
\item \texttt{GGA=PE}, the PBE Perdew-Burke-Ernzerhof functional.
\end{enumerate}

\indent All our samples were relaxed with \texttt{ISIF=3}, i.e. both atomic coordinates and cell dimensions were optimized. This does not, however, automatically rule out freezing of atomic coordinates and/or cell dimensions in local minima. Hence small increments as well as decrements of all cell dimensions have been performed with \texttt{ISIF=2} (fixed cell). Experimental lattice constants have been used only for~initialization of the relaxation procedure. When the $\mathrm{(N\times H)}$-polytypes were initialized with~increased or decreased volumes in an \texttt{ISIF=3}-type relaxation procedure, they spontaneously recovered their original volume. \\
\indent In addition, molecular dynamics runs accelerated by Machine-Learning Force Fields at~$1000\:\mathrm{K}$ lasting $\sim 10^{5}$ steps, $8\:\mathrm{fs}$ each, have been performed in \texttt{VASP} to further test whether the $\mathrm{(N\times H)}$-polytypes got stuck in local minima, but the corresponding energy decrease amounted merely to~$1-2\:\mathrm{meV/at.}$ and did not reduce the distance from the hull considerably. \\

\subsection{\label{meth:struct}Structures}

\indent Candidate structures for the La-O-W convex hull construction were mainly taken from the ICSD -- Inorganic Crystal Structure Database, \cite{icsd1}; the Materials Project (MP): database version v2025.09.25, \cite{MP}; Pearson's Handbook, \cite{Pearson}; and from~numerous articles dedicated to the topic. All the structures are minutely described in \cite{eva}. Our sample database is restricted to pure \emph{crystalline} phases. Symmetry of all the structures was determined \emph{via} the program \texttt{findsym}, \cite{findsym}. Relaxations were performed with symmetry switched off completely (\texttt{ISYM = -1}) to find the optimal ground state (GS) symmetry. All the structures were visualized by~the VESTA program,~\cite{vesta}.


\section{\label{results}Results}

\begin{figure*}[hbt!]
	\includegraphics[width = \textwidth]{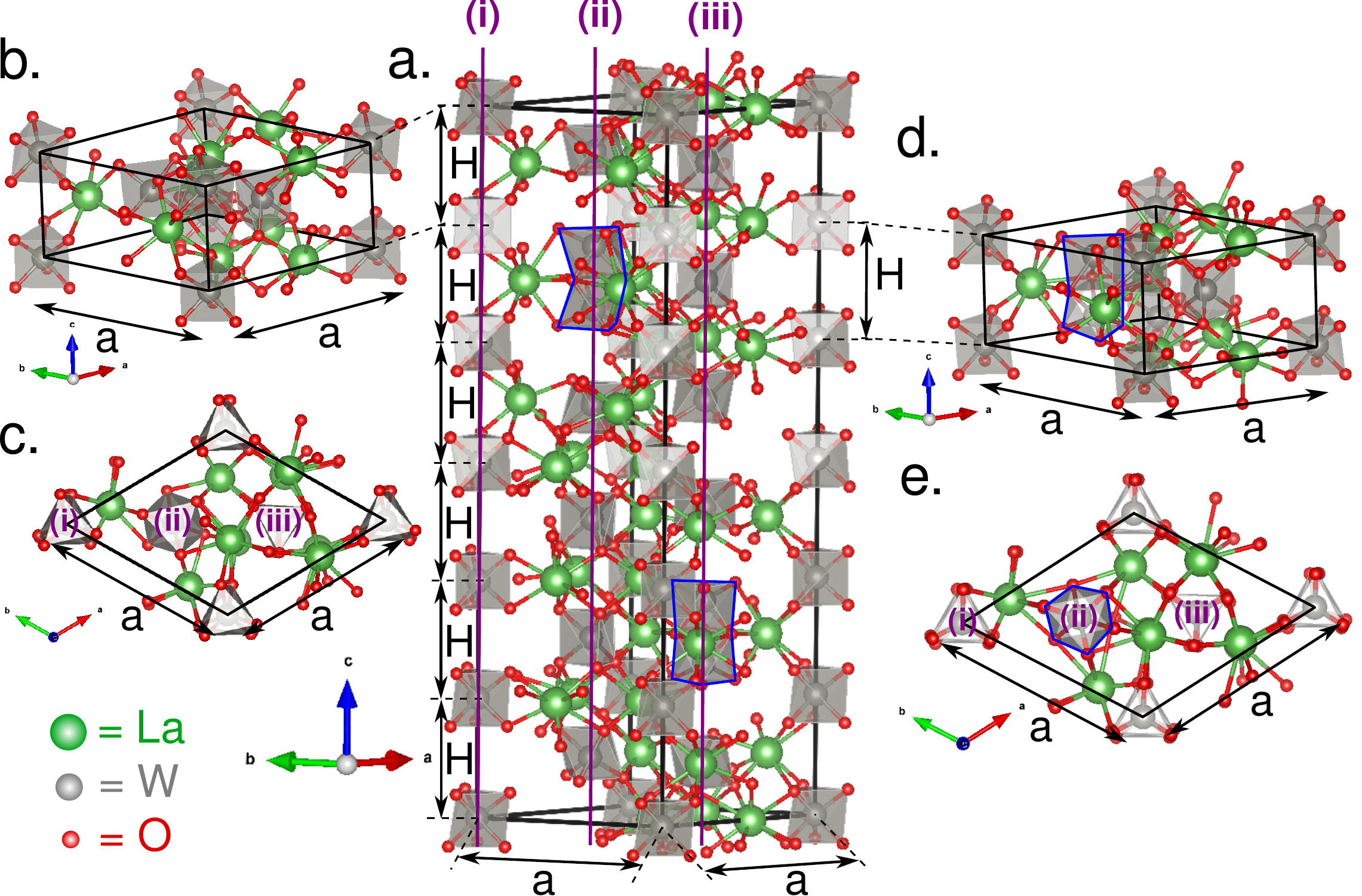}
	\caption{a.:~Unit cell of the most stable 6H-polytype of \ce{La18W10O57}.hP170 tungstate, constructed according to experimental data of N. E. Novikova, V. I. Voronkova et al., \cite{5H}, and the work of M.-H. Chambrier et al., \cite{la18w10o57}, consisting of 6 hexagonal basic blocks (H-blocks) of~height $H \approx 5.46\:$\AA, $c = 6H \approx 32.8\:$\AA, $a = b \approx 9.045\:$\AA. b.:~H-block without touching \ce{WO6} octahedra or trigonal prisms. c.:~Top view of the H-block from~b. d.:~H-block with a couple of~touching \ce{WO6} octahedra highlighted by \textcolor{blue}{\textbf{a solid blue line}}, sharing a~common triangular face. e.:~Top view of the H-block from~d. \textcolor{violet}{\textbf{(i)}}, \textcolor{violet}{\textbf{(ii)}} and \textcolor{violet}{\textbf{(iii) solid violet lines}} designate \textcolor{violet}{\textbf{a trio of W columns}} mentioned in the main text. \textbf{The hexagonal unit cell} is delimited by \textbf{a thick black line}; the main hexagonal crystallographic axes are represented by~vectors $\vec{a}$, $\vec{b}$ and $\vec{c}$.}
	\label{6H-fig}
\end{figure*}

\indent Non-stoichiometric hexagonal tungstates with composition close to \ce{La2WO6} are known as 4H-, 5H-, 6H- and 7H-polytypes, jointly $\mathrm{(N\times H)}$-\emph{polytypes} (i.e. structural variants), composed of N basic hexagonal building blocks of~height H, hereafter H-blocks, \autoref{6H-fig}. \\
\indent H-blocks stack upon each other along the hexagonal $c$ axis sharing three columns of \ce{WO6} units denoted as (i), (ii) and (iii) in \autoref{6H-fig}: column (i) of~trigonal prisms passes through the vertices of the hexagonal unit cell. The remaining two columns (ii), (iii) go through the centers of equilateral triangles of the unit cell base. All W atoms within the H-blocks are coordinated either by \ce{WO6} octahedra or by trigonal prisms aligned along the three columns, absorbing all O atoms present in the structure. In other words, after the removal of all \ce{WO6} units only the pure La scaffold remains. Both isolated and twinned \ce{WO6} units in the H-block interior are centered in the $c$ direction as well, positioned H/2 above the H-block base. As shall be discussed later, the substantial difference between W column (i) and the pair of W columns (ii), (iii) resides in the relative shift H/2 along the $c$ direction; the columns (ii), (iii) can equally well pass through the cell vertices, see \autoref{vector}. \\
\indent Further, each H-block contains 6 La atoms in two 3-La equilateral rings: one of~them is located in~H-block base and one in its interior. Both 3-La rings are symmetrically disposed around a \ce{WO6} octahedron/prism or around the~shared \ce{O3} face of twinned \ce{WO6} units. The only significant difference between the H-blocks is the presence or absence of ``twinned'' \ce{WO6} units, sharing a triangular \ce{O3} face, \autoref{6H-fig}. Note that an H-block with touching \ce{WO6} contains \emph{four}, instead of 3 W atoms. Even though all H-blocks have the same structure except for the presence of touching \ce{WO6} octahedra/prisms in~some of them, the precise composition (stoichiometry) of the $\mathrm{(N\times H)}$-polytypes given by the density of touching \ce{WO6} polyhedra and structural modulation along the $c$ axis are still unknown. For a review of the current literature, consult paper~\cite{eva}. \\
\indent DFT enables us to resolve various contradictory experimental results available for the 5H- and 6H-polytypes, obtained by various groups \emph{via} diffraction studies. The results of the DFT structure refinement are demonstrated in~the present \autoref{results}. The set of all stable compounds made of elements La, O, W at $T = 0\:\mathrm{K}$, $p = 0\:\mathrm{kbar}$ forms the so-called \emph{convex envelope}, which shall be given in the forthcoming \autoref{discussion}. The convex hull is direct evidence of~the reasonability of the empirical rules behind the polytype construction, suggested in this work. \\

\indent Atomic arrangement of the 6H-polytype with well-defined \ce{La18W10O57} stoichiometry and two such structural units per~cell was given by the group of M.-H. Chambrier \& F. Goutenoire et al., \cite{la18w10o57}, and by N. E. Novikova et al., \cite{5H}. The two atomic configurations, however, differ considerably. While Novikova et al.'s candidate lies practically on the hull ($\Delta E \approx 8\:\mathrm{meV/at.}$ apart, \autoref{fig-laow}), Chambrier's 6H-polytype is as much as $50\:\mathrm{meV/at.}$ above it. \\
\indent The main difference lies in the number of~face-sharing \ce{WO6} units: whereas Novikova's 6H-polytype contains two \emph{couples} of twinned \ce{WO6} trigonal prisms separated in $c$ direction by 2H, \autoref{6H-fig}, Chambrier's 6H sample exhibits two \emph{triples} of face-sharing \ce{WO6} in distinct W columns, extending over 2H each and mutually separated by 1H, which turned out~to~be unacceptable. \emph{Inter alia}, it results in six O atoms outside the \ce{WO6} polyhedra (two more shared \ce{O3} faces of \ce{WO6} units), bound only to La, which we believe is \emph{not} a characteristic feature of this family of structures. The most important quantitative result of this investigation is that the energy penalty for the trio of O atoms outside the \ce{WO6} units, caused by \emph{excessive} touching \ce{WO6} units sharing an~\ce{O3} triangular face, amounts to~$\approx 25\:\mathrm{meV/at}$. Note that removal of all these ``superfluous'' O atoms almost doubled the original energy distance from the hull to~$\Delta E \approx 100\:\mathrm{meV/at.}$ \\

\indent Various candidates for the 5H-polytype with putative composition \ce{La30W17O96} were proposed by N. E. Novikova et~al.~\cite{5H}. Energetically the best of them, however, lies $41\:\mathrm{meV/at.}$ above the convex hull, \autoref{fig-laow}, which is suspiciously far, taking into account the huge amount of tungstates falling into an $\approx 20\:\mathrm{meV/at.}$-wide window. What is more, the large number of W sites with rather low occupancies may suggest \emph{averaging over disorder} or substantial obstacles hindering structure determination from this particular diffraction experiment. Note that there are no other experimental data on 5H- nor any other $\mathrm{(N\times H)}$-polytype, except for the 6H-polytype discussed above. \\
\indent The most probable reason, why Novikova's proposal for the 5H-polytype was incorrect, is the presence of two couples of~touching \ce{WO6} units within a single H-block, that is 5 W atoms inside a 1H-block. Similarly to the incorrect identification of 6H-polytype by M.-H. Chambrier et~al., the problem most likely lies in the mutual repulsion of twinned \ce{WO6} units at too short a distance, i.e. too high a~density of~\ce{WO6}. \\
\indent Other energetically even more expensive 5H-candidates proposed by Novikova et al. suffer from an uneven number of La atoms in~individual H-blocks; the general rule of 6 La atoms per H-block is not obeyed by their atomic configurations. There are H-block bases containing as much as 6 La atoms, whereas some H-blocks embrace just 3 La atoms. Note that 6 La atoms per H-block base are characteristic of yet \emph{hypothetical} compound \ce{La6WO12}.hR57 with an analogous structure, which has been experimentally reported only for \ce{Y6WO12} and \ce{Ho6WO12}, \cite{Ln6WO12,Y6WO12}, but which turned out to be stable according to the present DFT evaluation, \cite{eva}, see also \autoref{fig-laow}. It is also interesting to see that in the course of DFT relaxations starting from these uneven La distributions, they gradually approach the expected outcomes outlined below, so the DFT relaxations themselves show the potential to improve the insufficiently captured experimental data. \\

\begin{table}[h!]
{
\centering
\renewcommand\cellalign{lc}
\setcellgapes{3pt}\makegapedcells
{\normalsize{
\begin{tabular}{|c|F{0.95\textwidth}|}
\hline
no. & rule   \\
\hline
\hline
1. & All W atoms have six O nearest neighbours. \\
\hline
2. & \ce{WO6} units are aligned along columns (i), (ii), (iii) in the $c$ direction: (i) passing through the hexagonal cell vertices and (ii), (iii) running through the centers of the equilateral-triangle cell base, see \autoref{6H-fig}. \\
\hline
3. & All W atoms on column (i) are coordinated by~trigonal prisms; W atoms on columns (ii), (iii) are mostly octahedrally coordinated. \\
\hline
4. & W atoms in columns (ii), (iii) have the same vertical $c$ positions. Column (i) is vertically shifted by H/2 with respect to (ii), (iii), which can equally well pass through the cell vertices, see \autoref{vector}. \\
\hline
5. & All O atoms belong to the \ce{WO6} coordination polyhedra. \\
\hline
6. & Each H-block contains 6 La atoms in two symmetrical 3-La rings centered around W columns. One of~the two 3-La rings lies in the H-block base and the second surrounds either the interior \ce{WO6} or the shared \ce{O3} face of a twinned \ce{WO6}. \\
\hline
7. & Twinned \ce{WO6} units occupy only the two W columns (ii), (iii) sharing the same $c$ coordinates of W atoms. \\
\hline
8. & Twinned \ce{WO6} octahedra can share \ce{O3} faces only in pairs, not in triplets, quadruplets, etc. \\
\hline
9. & Twinned \ce{WO6} pairs in one, e.g. (ii), W column have to be separated by at least 2H blocks along the $c$ direction. \\
\hline
10. & A twinned \ce{WO6} pair in the other, e.g. (iii), interior W column has to be separated along the $c$ axis by at least 1H from the closest \ce{WO6} pair in column (ii). \\	
\hline
\end{tabular}}}
\caption{\label{ttt}{\normalsize{Simple empirical rules governing $\mathrm{(N\times H)}$-polytype construction.}}}}
\end{table}

\begin{figure*}[hbt!]
\includegraphics[width = \textwidth]{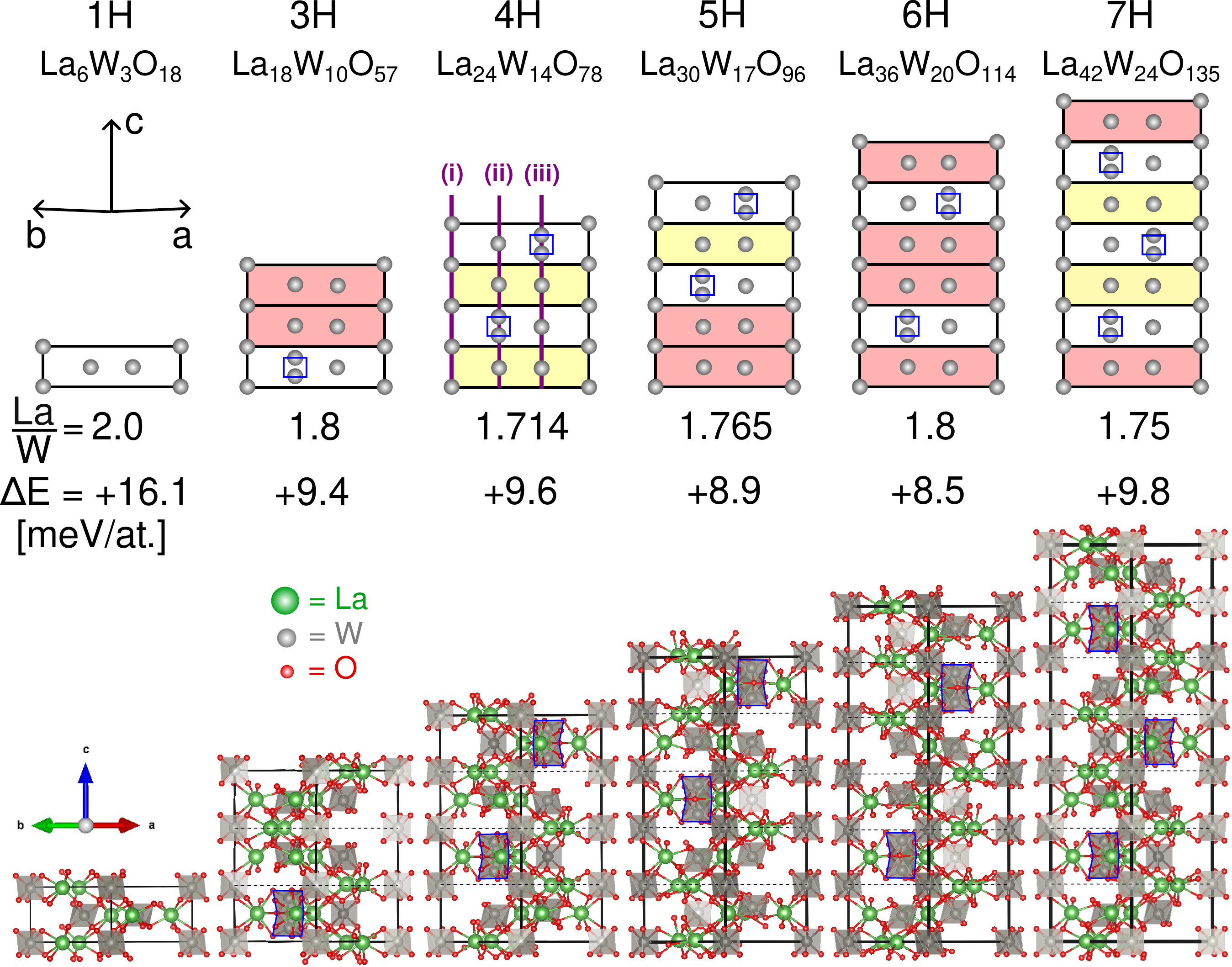}
\caption{\underline{Upper panel:} Schematic representation of $\mathrm{(N \times H)}$-polytypes (showing W atoms only). Below each polytype label its stoichiometry per periodic unit cell is stated. A single H-block (1H) is depicted as a black rectangle. \textcolor{violet}{\textbf{Three W columns}} are designated as \textcolor{violet}{\textbf{(i), (ii)}} and \textcolor{violet}{\textbf{(iii)}}, like in \autoref{6H-fig}. \textcolor{blue}{\textbf{``Glued'' \ce{WO6} units}}, sharing an \ce{O3} base, are highlighted by \textcolor{blue}{\textbf{a small blue rectangle}}. H-blocks shaded in~red stand for~(minimal) 2H-wide separation of H-blocks containing glued \ce{WO6} units in the \emph{same} W column. H-blocks shaded by~yellow stand for a minimal 1H-wide separation of H-blocks containing glued \ce{WO6} units in \emph{distinct} W columns. The main hexagonal crystallographic axes are visualized as a trio of vectors $\vec{a}$, $\vec{b}$ and $\vec{c}$. Below~each polytype, the ratio of the number of~atoms $\mathrm{La:W}$ is stated, together with the respective energy distance $\Delta E$ from the La-O-W convex hull, \autoref{fig-laow}, in~units of~meV/at. \underline{Lower panel:} Ball-and-Stick representation of $\mathrm{(N \times H)}$-polytypes above (the same view direction). Periodic unit cells are delimited by a solid black line; individual H-blocks are separated by thin dashed lines; \textcolor{blue}{\textbf{twinned \ce{WO6} units}} -- occupying exactly the same positions as in the scheme above -- are rimmed by~\textcolor{blue}{\textbf{a solid blue line}}.}
\label{schematic}
\end{figure*}

\begin{figure*}[hbt!]
\includegraphics[width = 1.0\textwidth]{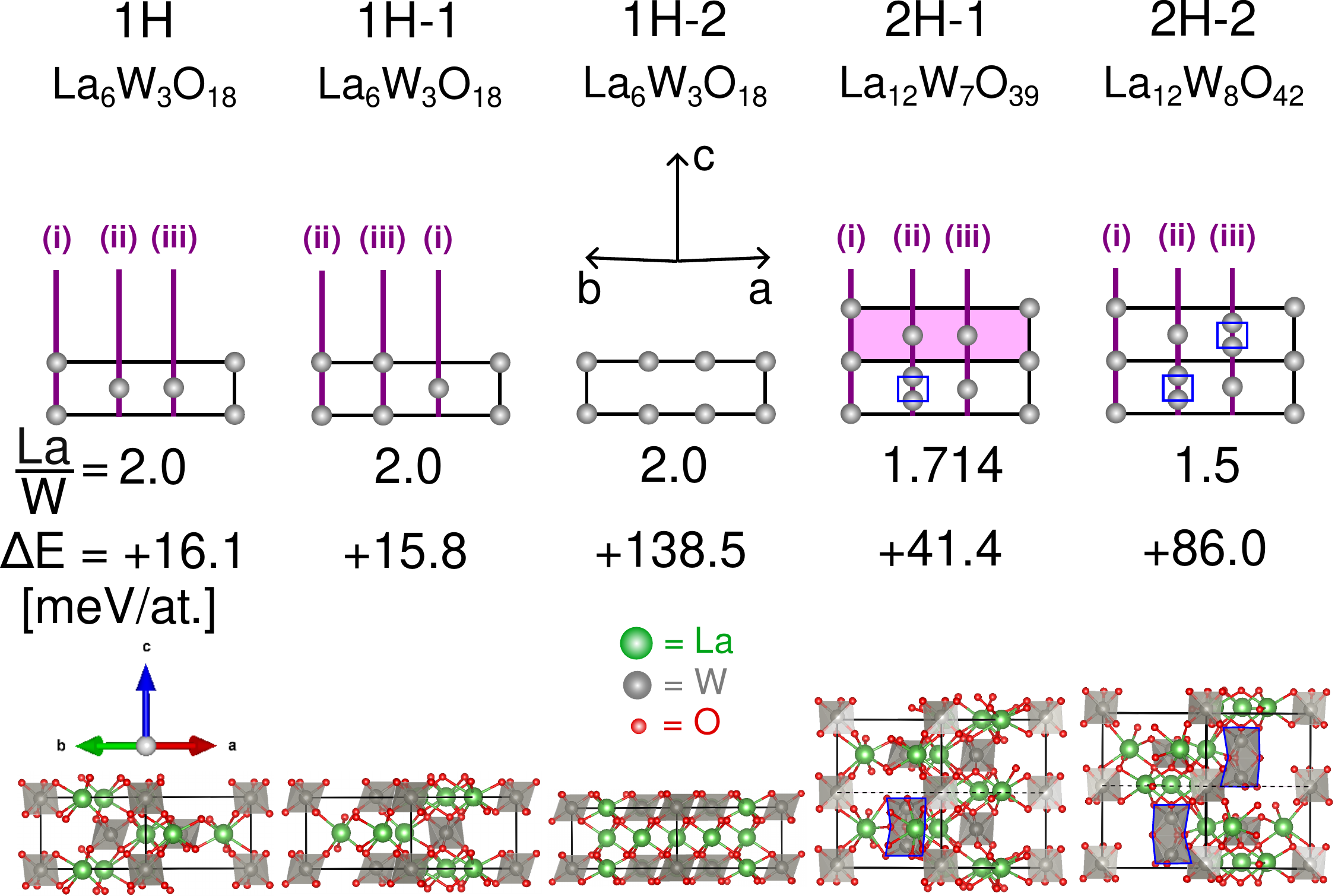}
\caption{Schematic representation of 1H- and 2H-polytypes. The two versions of the 1H-polytype: 1H and 1H-1 without twinned \ce{WO6} units are geometrically equivalent; they are mutually related \emph{via} a horizontal translation in the $ab$ plane, see \autoref{vector}. The third version, 1H-2, is obviously not experimentally viable. Both options of the 2H-polytype are clearly unstable ($\Delta E > 40\:\mathrm{meV/at.}$), owing to the too short a~distance between glued \ce{WO6}. The logic behind this representation is the same as in~\autoref{schematic}.}
\label{bad}
\end{figure*}

\begin{figure*}[hbt!]
\includegraphics[width = 0.45\textwidth]{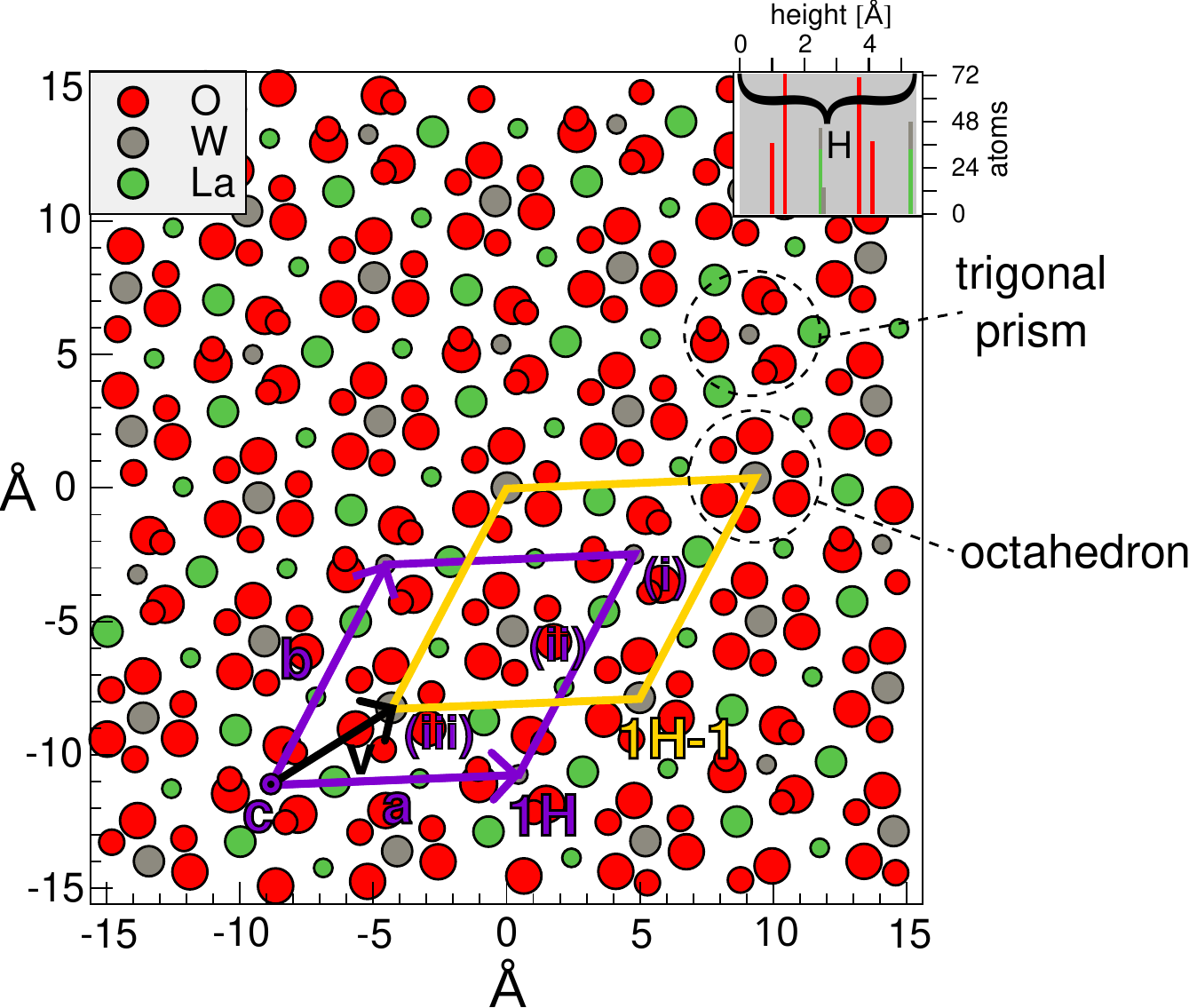}
\caption{Two equivalent choices of the $\mathrm{(N\times H)}$-polytype unit cell, delimited by orange and violet solid lines, corresponding to~geometrically equivalent 1H-1 and 1H structures depicted in \autoref{bad}, respectively. The two cells are mutually shifted by~$\vec{v} = (1/3)\vec{a} + (1/3)\vec{b}$ in the horizontal $ab$ plane. Atomic sizes scale with the distance from the observer along the hexagonal $c$ axis. Both cells contain two W columns (ii), (iii) with W atoms occupying the same $c$ positions (passing through the bigger W atoms) and the third W column (i) shifted along the $c$ axis by H/2 (passing through the smaller W atoms), where H stands for the height of the H-block. Note that W atoms in columns (i) are surrounded by O in trigonal prisms, whereas W atoms in~columns (ii), (iii) exhibit an octahedral O environment. The histogram in the upper right corner shows the number of atoms of~each kind in~the depicted $a-b$ window per vertical $c$ coordinate.}
\label{vector}
\end{figure*}

\indent The only DFT-verified experimentally observed 6H-polytype, as well as many unsuccessful attempts to~describe 6H- and 5H-polytypes, inspired us to formulate the following \emph{empirical} rules governing the $\mathrm{(N\times H)}$-polytype construction, N$\,\geq 3$, \autoref{ttt}. It is the very last rule (no.~10) that may yield 4H-, 5H- and 7H-polytypes with periodicity other than $\mathrm{N\times 3H}$ in~the $c$ direction. $\mathrm{(N\times H)}$-polytypes constructed according to rules 1--10 are schematically depicted in~\autoref{schematic}, from~which it is clear that the polytypes cover the interval of atomic ratios $\mathrm{La:W} \in [1.714,\,2]$ continuously, since an arbitrarily small concentration of glued \ce{WO6} units \emph{is} in compliance with the building principles 1--10. \\
\indent Let us emphasize that rules 1--10 in \autoref{ttt} are based entirely on direct experimental observations, mainly neutron and X-ray diffraction fits, and a thorough theoretical analysis calls for further verification of some of them, \autoref{t2}. The last two points/suggestions in \autoref{t2}, when concerning the whole structure, should be called ``2H-polytypes'', which are forbidden in our present models, owing to the repulsive interactions between glued \ce{WO6} units in the same or distinct interior W columns, see \autoref{bad}. In the subsequent studies presented in \autoref{discussion}, we will check assumptions from \autoref{t2} by explicit construction of 2H-, 3H-, 4H-, 5H- and 7H-polytypes, most of which have been observed experimentally (except for 2H- and 3H-polytypes). \\
\indent The first corroboration of the above mentioned hypotheses 1--10 is that the 1H block \emph{without} touching \ce{WO6} polyhedra periodically stacked in the $c$ direction with the stoichiometric composition \ce{La2WO6} has admissible energy, situated merely $\Delta E \approx 16\:\mathrm{meV/at.}$ above the stable \ce{La2WO6}.oP36 variant, \autoref{fig-laow}. This implies that the chief contribution to the internal energy comes from the interaction between twinned \ce{WO6} units, which seem to repel each other at~small distances ($\mathrm{\leq\,1H}$ or so). Nevertheless, since \ce{La18W10O57}-6H is twice as close to the convex hull ($\Delta E = 8\:\mathrm{meV/at.}$) compared with~\ce{La2WO6}-1H ($\Delta E = 16\:\mathrm{meV/at.}$), it follows that glued \ce{WO6} units indeed \emph{have} some stabilizing effect on~the $\mathrm{(N \times H)}$-polytypes with~diverse $c$-periodicity. Therefore it makes sense to search for the highest permissible density of glued \ce{WO6} pairs, leading to~the distinct stoichiometries and varying $c$-periodicity. \\

\vspace{0.15cm}

\begin{table}[h!]
{
\centering
\renewcommand\cellalign{lc}
\setcellgapes{3pt}\makegapedcells
{\normalsize{
\begin{tabular}{|c|F{0.95\textwidth}|}
\hline
no. & energy penalty of a \ce{WO6} arrangement to be computed \\
\hline
\hline
1. & twinned \ce{WO6} units in a W column (i), shifted by H/2 with respect to W columns (ii), (iii), \\
\hline
2. & different symmetry-breaking options caused by relative shifts of W columns in the $c$ direction, e.g. those characteristic of \ce{La2WO6}.oP36 or $\alpha$-\ce{La2WO6}, see \autoref{discussion}, \\
\hline
3. & twinned \ce{WO6} units in the same W column mutually separated by a gap smaller than 2H (i.e. 1H or even 0H), \\
\hline
4. & twinned \ce{WO6} units in distinct W columns situated in neighboring H-blocks (mutually separated by 0H). \\
\hline
\end{tabular}}}
\caption{\label{t2}{\normalsize{\ce{WO6} arrangements which have to be checked in order to verify the simple empirical rules from \autoref{ttt}.}}}}
\end{table}

\vspace{0.0cm}

\begin{figure*}[hbt!]
\includegraphics[width = 1.0\textwidth]{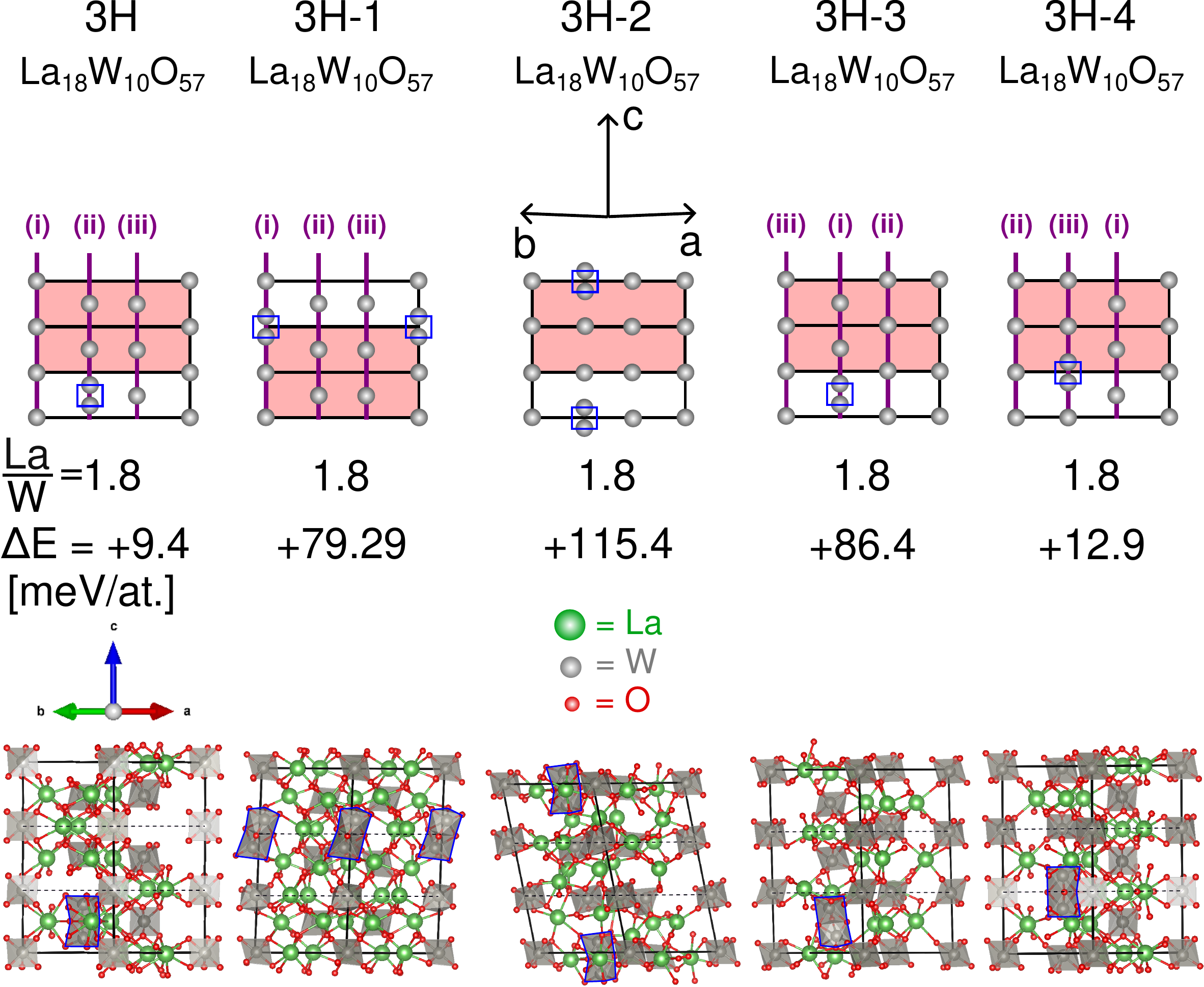}
\caption{Different variations of the 3H-polytype produced by relative shifts of W columns along the hexagonal $c$ axis. 3H and 3H-4 variants are geometrically equivalent, mutually related via a horizontal translation in the $ab$ plane, see \autoref{vector}, hence they have approximately the same energy. 3H-2, like 1H-2 above (see \autoref{bad}), is obviously not experimentally viable. 3H-1 and 3H-3 (also mutually equivalent) contain twinned \ce{WO6} units in the forbidden W column (i), hence the high internal energy.}
\label{bad1}
\end{figure*}


\section{\label{discussion}Discussion}

\subsection{\label{La-O-W}La-O-W Convex Hull}

\begin{figure*}[hbt!]
\includegraphics[width = \textwidth]{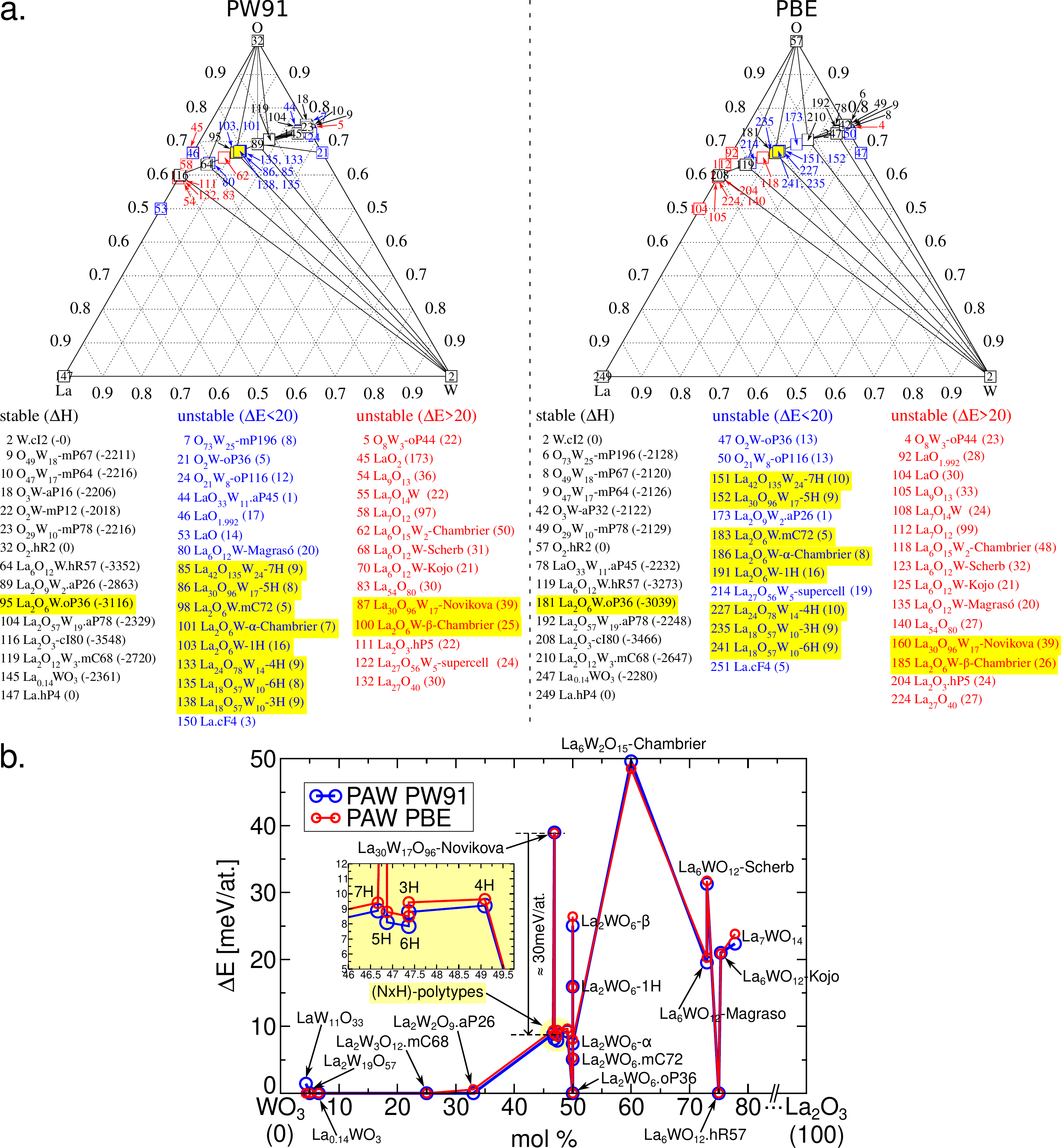}
\caption{\underline{Upper panel a.:} Ternary La-O-W convex hull at $p=0\:\mathrm{kbar}$, $T=0\:\mathrm{K}$. Ternary compounds belonging to the \ce{La2WO6} family are \hlc[yellow]{highlighted in yellow}. Both panels yield approximately the same results, as only the type of DFT functional differs: \texttt{GGA=91} on~LHS vs. \texttt{GGA=PE} on RHS. The internal-energy distance $\Delta E$ from the envelope is stated after the compound's Pearson symbol in~parenthesis in~units of $\mathrm{meV/at}$. Compounds are classified into three categories: \textbf{stable (black)}, \textcolor{blue}{\textbf{slightly unstable (blue, $\mathbf{\Delta E \leq 20\:\mathrm{\mathbf{meV/at.}}}$)}} and \textcolor{red}{\textbf{unstable (red, $\mathbf{\Delta E > 20\:\mathrm{\mathbf{meV/at.}}}$)}}. For a stable XY compound, the number $\Delta H$ in~parenthesis has the meaning of the per-atom energy distance from a \emph{hypothetical} convex envelope constructed \emph{after} XY removal from~the database. The convex envelope was created using the program \texttt{qhull}, \cite{qhull}. \underline{Lower panel b.:} Projection of~the ternary convex hull above into \ce{WO3}--\ce{La2O3} line segment. All the investigated La tungstates ``LaWO'' were decomposed according to~the chemical reaction: ``\ce{LaWO}'' $\mapsto$ x\ce{La2O3} + y\ce{WO3} + z\ce{O2}, x + y = 1 ($\mathrm{\Rightarrow\,\,x = 1/(2W/La + 1)}$, where W and La stand for~the respective number of atoms in ``LaWO''), since the amount of La and W is given by the quantity of the two powder precursors, whereas the residual O can be interchanged with the atmosphere. The DFT-computed internal-energy reduction of~5H-polytype (\ce{La30W17O96}) \hlc[lightye]{(shaded by yellow)} with~respect to the results of Novikova et~al. equals $\approx 30\:\mathrm{meV/at}$.}
\label{fig-laow}
\end{figure*}

\begin{figure*}[hbt!]
	\includegraphics[width = 0.8\textwidth]{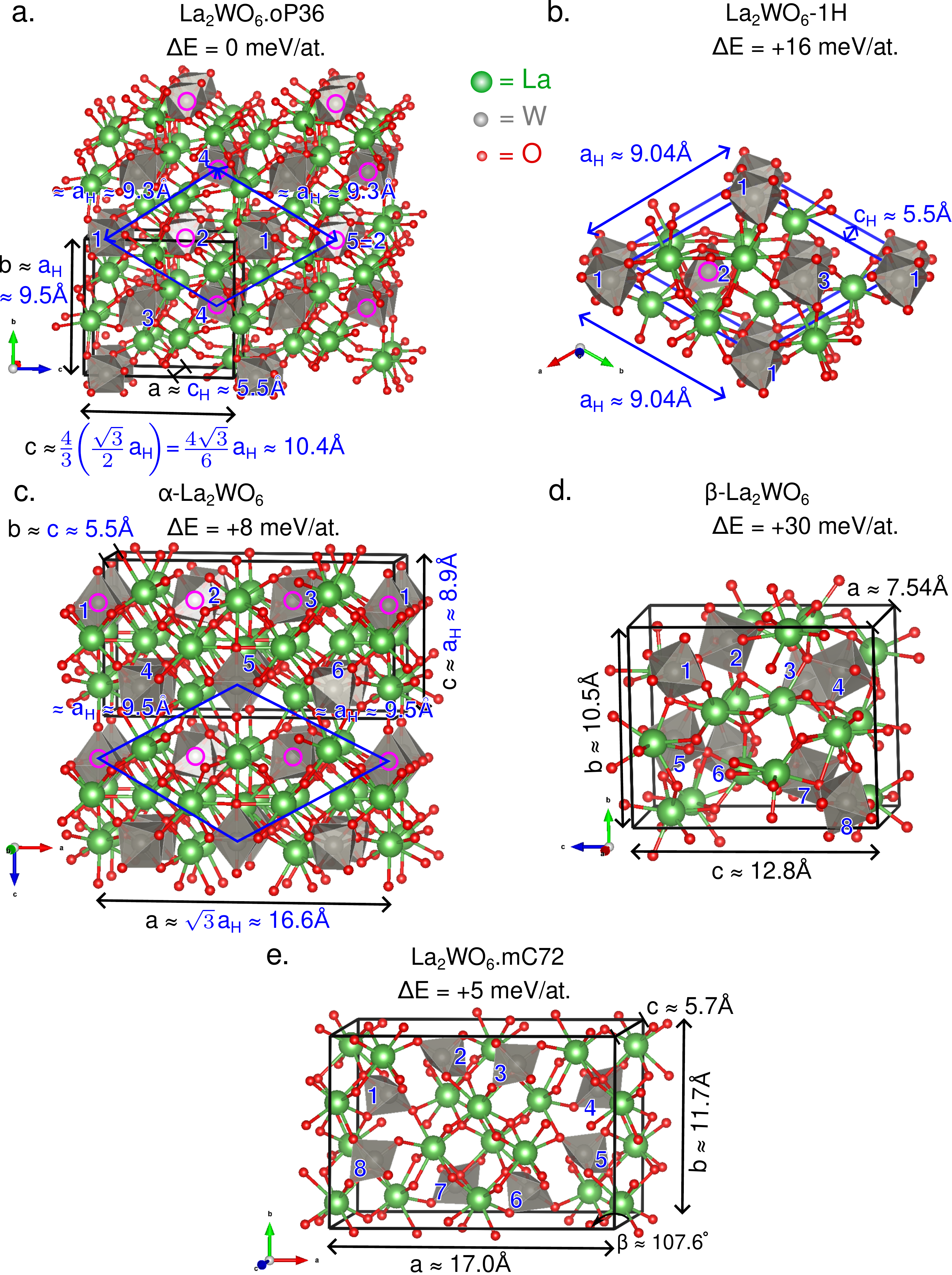}
	\caption{Quintet of structures with the \ce{La2WO6} composition. a.:~Unit cell of the most stable \ce{La2WO6}.oP36 obtained by~Gd$\,\mapsto\,$La substitution in \ce{Gd2WO6}, \cite{gd2wo6}, 4W/cell (atom 5 is a periodic image of~atom 2). b.: 1H hexagonal \ce{La2WO6} block without~glued \ce{WO6} octahedra from~\autoref{6H-fig}, panel~b, 3W/cell. c.: High-temperature $\alpha$-\ce{La2WO6} from diffraction experiments of M.-H. Chambrier et~al. possessing 6W/cell, \cite{a-la2wo6}. d.:~Low-temperature $\beta$-\ce{La2WO6} from diffraction experiments of M.-H. Chambrier et~al., 8W/cell, \cite{b-la2wo6}. e.:~Unit cell of \ce{La2WO6}.mC72, a \emph{hypothetical} structure retrieved from MP (mp-770889), \cite{MP}, 8W/cell, which has been prepared experimentally as \ce{Sm2MoO6} (ICSD 4186), \cite{sm2moo6}. The hexagonal 1H unit cell from panel b. is delimited also in panels a. and c. by~\textcolor{blue}{\textbf{solid blue lines}} for~comparison. The unit cell relevant to each panel is displayed as~\textbf{a solid black line}. W atomic columns shifted by $1/2$ of the cell vector perpendicular to~the hexagonal plane with respect to the rest of~the W columns (whose W atoms lie on the cell foremost face with~hexagonal symmetry) are highlighted by \textcolor{magenta}{\textbf{pink circles}}. The relation between the black and blue unit cells is indicated (except for~panels d., e.) and described in the main text. The main crystallographic axes are represented by a trio of~vectors $\vec{a}$, $\vec{b}$ and $\vec{c}$. The internal-energy distance from the convex hull, $\Delta E$, is stated below the polytype formula. Distinct W positions in~the unit cell are marked by~\textcolor{blue}{blue figures} in each case.}
	\label{1H-fig}
\end{figure*}

\indent La-O-W ternary convex hull at $p=0\:\mathrm{kbar}$, $T=0\:\mathrm{K}$ is presented here, \autoref{fig-laow}, where all the competing compounds are classified into three groups by the internal-energy distance $\Delta E$ from the envelope stated after the compound's Pearson symbol in~parenthesis in~units of $\mathrm{meV/at.}$: \textbf{1. stable phases (black)} forming the envelope, \textcolor{blue}{\textbf{2. slightly unstable or metastable phases (blue) up~to~$\mathbf{20\:\mathrm{\mathbf{meV/at.}}}$ from~the~envelope}} and \textcolor{red}{\textbf{3. highly unstable phases (red) more than $\mathbf{20\:\mathrm{\mathbf{meV/at.}}}$ above the hull}}. Note that the threshold $20\:\mathrm{meV/at.}$ was chosen (somewhat arbitrarily) because it corresponds to $\approx 230\:\mathrm{K}$ in $k_{\mathrm{B}}T$ units and hence all the compounds from~the second category labeled ``slightly unstable'' are in~principle accessible to experimental preparation. The convex envelope, however, concerns the conditions of~true thermodynamic equilibrium, which is not usually relevant to experimentally known oxides, being prepared under high temperatures and quench rates \emph{via} specific chemical reactions from the respective precursors. \\ 
\indent The number $\Delta H$ in~parenthesis after the \emph{stable} compound's XY Pearson symbol has the meaning of the per-atom energy distance from the hypothetical convex envelope constructed \emph{after} XY removal from~the database. \\
\indent According to our DFT computations, the still \emph{hypothetical} orthorhombic stoichiometric \ce{La2WO6}.oP36 sample ($a=5.494\:$\AA, $b=9.488\:$\AA, $c=10.43\:$\AA) obtained \emph{via}~substitution Gd$\,\mapsto\,$La in \ce{Gd2WO6}, \cite{gd2wo6}, see \autoref{1H-fig}, was $8\:\mathrm{meV/at.}$ lower than the high-temperature experimentally observed $\alpha$-\ce{La2WO6}, which in turn was -- surprisingly -- $\approx 20\:\mathrm{meV/at.}$ \emph{below} the low-temperature experimentally observed $\beta$-\ce{La2WO6}, cf.~paper \cite{b-la2wo6}, so the experimental prediction of~M.-H. Chambrier et al. of the GS \ce{La2WO6} atomic configuration from~\emph{ab~initio} diffraction experiments was again $\approx 30\:\mathrm{meV/at.}$ above the DFT GS. Hence we estimate the reliability of such experimental predictions of La tungstate structure from~\emph{ab~initio} diffraction fitting to be $\approx 30-50\:\mathrm{meV/at.}$, which is not an insignificant uncertainty, which DFT \emph{can} eliminate. \\
\indent The most stable \ce{La2WO6} polymorph, \ce{La2WO6}.oP36, is structurally closely related to the 1H-block without touching \ce{WO6} units from the preceding \autoref{results} ($a = b = 9.310\:$\AA, $c=5.463\:$\AA), \autoref{6H-fig}, panel b. Within its orthorhombic unit cell the hexagonal unit cell of the 1H-block is clearly identified, \autoref{1H-fig}. The lower symmetry of~oP36 (space group no.~19) compared to~that of the H-block (space group no.~150) is a consequence of the shifts of W columns along the hexagonal $c$ axis, corresponding to~the $a$ axis of the resulting orthorhombic \ce{La2WO6}.oP36 phase. Namely, atom ``5'' (the periodic image of atom ``2'') and atom ``4'' in \ce{La2WO6}.oP36, \autoref{1H-fig}a, are both shifted by $a/2$ with respect to atom ``1'' of \ce{La2WO6}-1H, \autoref{1H-fig}b; they are marked by pink circles. Both atoms ``5'' and ``4'' correspond to the same atomic position ``1'' in \ce{La2WO6}-1H. Atom ``2'' in~\ce{La2WO6}.oP36 corresponds to~the homonymous site with the same coordinates in \ce{La2WO6}-1H. The periodic image ``1'' \emph{inside} the blue hexagonal unit cell delimited in \autoref{1H-fig}a corresponds to a different site ``3'' (i.e. different from ``1'') in~\autoref{1H-fig}b with the same coordinates. \\
\indent Note that the thickness of the H-block in the $c$ direction, $c\approx 5.46\:$\AA, exactly equals the periodicity of~\ce{La2WO6}.oP36 in the $a$ direction. Because the area of the unit-cell base perpendicular to the unit-cell edge with the equal length $\approx 5.46\:$\AA\, is (4/3) times larger for the \ce{La2WO6}.oP36 compared to \ce{La2WO6}-1H, oP36 contains 4 W atoms per cell instead of 3 W atoms per cell of the 1H-polytype. Therefore the atomic density of both structures is exactly the same. \\
\indent Moreover, in both structures all O atoms are contained within \ce{WO6} octahedra, as in the $\mathrm{(N\times H)}$-polytypes. Energetically, the two structures are only $16\:\mathrm{meV/at.}$ apart, \autoref{fig-laow}, which is comparable to~the magnitude of~the thermal fluctuations at room temperature. Together with the observation that 3H-, 4H-, 5H-, 6H- and 7H-polytypes lie practically on the hull, $\Delta E \approx 8\:\mathrm{meV/at.}$, \autoref{fig-laow}, it further confirms that the construction of~$\mathrm{(N\times H)}$-polytypes based on the modulated stacking of H-blocks suggested in the present work is plausible. It follows that inclusion of~a~small quantity of \ce{WO3} molecules into the \ce{La2WO6}-1H structure has an overall stabilizing effect, lowering its distance from the hull by $0.33-0.6\:\mathrm{eV}$ per one additional \ce{WO3} molecule. \\
\indent Direct evidence of the shortest possible distances between mutually repelling twinned \ce{WO6} units in the interior W columns (ii), (iii), cf. \autoref{schematic}, was given by the explicit construction of two variants of the \emph{hypothetical} 2H-polytype, \autoref{bad}. The first variant, \ce{La12W7O39}, contains all glued \ce{WO6} units in the same (ii) W column, separated by a 1H distance, which was assumed to be forbidden (too small). This assumption was verified by the respective distance above the convex envelope: $\Delta E \approx 40\:\mathrm{meV/at.}$, implying instability of the thus-constructed 2H-polytype. Compared to the 1H-polytype, it brings an energy penalty $\approx +1.4\:\mathrm{eV}$ per one additional \ce{WO3} molecule. The second variant, \ce{La12W8O42}, accommodating twinned \ce{WO6} in each H-block, which alternate in (ii) and (iii) W columns along the $c$ axis, exhibits even much worse energetics than \ce{La12W7O39}, incurring an energy penalty $\approx +1.9\:\mathrm{eV}$ per one additional \ce{WO3} molecule. In conclusion, the width of the gap between glued \ce{WO6} units has to be at least that given by yellow and red shaded panels displayed in \autoref{schematic}. \\
\indent As to the symmetry breaking caused by~shifts of selected W columns along the hexagonal $c$ axis, it was verified that all structures with two W columns (ii), (iii) shifted by H/2 with respect to the third W column (i) along the $c$ axis are equivalent and related by horizontal translation by vector $\vec{v} = (1/3)\vec{a} + (1/3)\vec{b}$ in the a-b plane, \autoref{vector}. Namely, on the example of the 1H- and 3H-polytype it has been manifested that alignment of all three W columns results in a considerable energetic penalty ($\Delta E > 100\:\mathrm{meV/at.}$) and that twinned \ce{WO6} units can only be created in columns (ii), (iii), see Figures \ref{bad}, \ref{bad1}. \\
\indent High-temperature $\alpha$-\ce{La2WO6} ($a=16.67\:$\AA, $b=5.558\:$\AA, $c=8.920\:$\AA, space group no.~31) determined by M.-H. Chambrier et al. from diffraction experiments \cite{a-la2wo6} is also closely related to the GS allotrope \ce{La2WO6}.oP36, which is reflected in its low distance from the hull ($\Delta E = 8\:\mathrm{meV/at.}$), \autoref{1H-fig}c. Here every second row of W columns along the $c$ axis is shifted by $b/2$ with respect to the others (marked by~pink circles), resulting in a staggered W configuration. Note that rows of W columns shifted in \autoref{1H-fig}c (alpha-La2WO6) are perpendicular to the rows shifted in \autoref{1H-fig}a (La2WO6.oP36). The periodicity in the $b$ direction ($\approx 5.5\:$\AA) agrees with the~1H periodicity along the hexagonal axis $c$. All O atoms are again consumed by \ce{WO6} octahedra. The presence of six W atoms per unit cell is also in compliance with the unit cell volume, which is roughly (3/2) times that of~the \ce{La2WO6}.oP36 variant with 4W/cell. \\
\indent Paradoxically, the low-temperature GS candidate of M.-H. Chambrier et~al., $\beta$-\ce{La2WO6} ($a=7.538\:$\AA, $b=10.52\:$\AA, $c=12.84\:$\AA, space group no.~19), \cite{b-la2wo6}, exceeds in energy that of the high-temperature form $\alpha$-\ce{La2WO6} from the same experiment. Based on our above mentioned structural considerations and DFT calculations, $\beta$-\ce{La2WO6} does not represent the GS atomic configuration with \ce{La2WO6} stoichiometry, lying as much as $\Delta E \approx 30\:\mathrm{meV/at.}$ far from the convex envelope. This particular structure has no obvious relationship to~any of the preceding: \ce{La2WO6}.oP36, \ce{La2WO6}-1H, $\alpha$-\ce{La2WO6} or \ce{La18W10O57}. Its most striking feature are 4 pairs of vertex-sharing \ce{WO6} octahedra per unit cell, resulting in 4 O atoms outside the \ce{WO6}, which in turn no longer form columns nor any hexagonal arrangement. In~this case, the disposition of \ce{WO6} octahedra is rather chaotic, \autoref{1H-fig}d. \\
\indent All the above mentioned low-energy members of the \ce{La2WO6} family suggest that W atoms enforce octahedral or prismatic \ce{WO6} nearest-neighbour coordination shell. This was further corroborated by the \ce{La6WO12} family of~structures, where configurations with \ce{WO5} pyramidal environments seem to bring an energy penalty, \cite{eva}. We have found, however, that the still \emph{hypothetical} \ce{La2WO6}.mC72 isostructural with experimentally prepared \ce{Sm2MoO6}, \cite{sm2moo6}, with all W atoms surrounded by \ce{WO5} trigonal bipyramids, \autoref{1H-fig}e, is situated only $\Delta E \approx 5\:\mathrm{meV/at.} \cong 60\:\mathrm{K}$ above the convex hull, \autoref{fig-laow}. \\
\indent \ce{La2WO6}.mC72, belonging to space group no.~15 ($a=17.04\:$\AA, $b=11.70\:$\AA, $c=5.669\:$\AA, $\beta \approx 107.6^{\circ}$), is markedly distinct from all the other members of the \ce{La2WO6} family mentioned in the present contribution. The unit cell comprises eight \ce{WO5} trigonal bipyramids forming 8-rings in the a-b plane, highly puckered in the $c$-direction, \autoref{1H-fig}e. The nearest-neighbour W-W distance is only $\approx 4\:$\AA. All La atoms within this structure occupy centers of~distorted \ce{LaO8} coordination cubes. \\

\indent Direct construction of the La-O-W convex hull by DFT, \autoref{fig-laow}, implies that in the PW91 setup all $\mathrm{(N\times H)}$-polytypes, $\mathrm{N} \in \{3,\,4,\,5,\,6,\,7\}$, with $\mathrm{La:W} \in [1.714,\,2]$ are unstable against decomposition into \ce{La2WO6} and \ce{La2W2O9}, cf. \autoref{t3}. This is in compliance with experimental observations of~Chambrier et al., \cite{la18w10o57}, and Abeysinghe et al., \cite{ce18w10o57}, stating that a larger amount of \ce{WO3} (than corresponds to~the polytype family) leads to~\ce{La2W2O9} as an impurity, whereas a~smaller amount of~\ce{WO3} yields~\ce{La2WO6} as an impurity phase. On the contrary, the PBE functional gives decomposition into~\ce{La2WO6} and \ce{La2W3O12}, but the energy difference from~the \ce{La2WO6} + \ce{La2W2O9} combination is smaller than that between PW91 and PBE and is of~the order of~1--2\%, since both ternary compounds \ce{La2W2O9}, \ce{La2W3O12} lie on the convex hull. It should be stressed, however, that the decomposition of the $(\mathrm{N\times H})$-polytypes is also governed by their interfacial energy with the candidate phases: \ce{La2W2O9} or \ce{La2W3O12}. \\
\indent Therefore we assume that these slightly \emph{metastable} polytypes are stabilized by their composition: they are formed in a slight surplus of W atoms compared with the stoichiometric \ce{La2WO6}.oP36 and can absorb in principle an arbitrarily small amount of excess W. It seems that this is the way nature deals with a tiny surplus of \ce{WO3} species. One of the reasons for the metastable polytype creation could be the avoidance of interface formation between \ce{La2WO6} and stable phases with~higher \ce{WO3} content, such as \ce{La2W2O9} or \ce{La2W3O12}. \\
\indent Note that \emph{all} $(\mathrm{N\times H})$-polytypes have $\Delta E$ very close to~$0\:\mathrm{meV/atom}$, namely $\Delta E \approx 8.5\:\mathrm{meV/at.} \cong 100\:\mathrm{K}$ comparable with thermal fluctuations at room temperature. A~few $\mathrm{meV/atom}$ instability that we encounter certainly means that our optimized structures are absolutely plausible. The central finding of our study is that in this family of~\ce{La2WO6}-related oxides, extra W atoms forming twinned \ce{W2O9} octahedra practically do not affect the value of $\Delta E$, as suggested by DFT. In the present work we do not take into account finite temperatures by including phonon free energies, so the impact of~vibrational entropy cannot be ruled out. \\
\indent In the limit of a trace amount of twinned \ce{WO6}, they are randomly distributed in the structure, resulting in effective 1H-periodicity. On the other hand, for sufficiently high a W density in the compound, discrete periodicities dominate (3H--7H), because the more glued \ce{WO6} units, the lower the configurational entropy. When the concentration of W atoms exceeds some critical threshold, the closest W-rich thermodynamically stable phase -- \ce{La2W2O9} -- starts to segregate. Note that the 3H-polytype with exactly the same stoichiometry as the 6H-polytype (\ce{La18W10O57}) has not been experimentally observed yet, \cite{la18w10o57}. This may be another manifestation of the repulsion between glued \ce{WO6} units. \\

\begin{table}[h!]
\centering
\small{
\begin{tabular}{|l|l|r|l|l|}
\hline
\rowcolor{olive9}$\mathrm{N\!\!\times\!\!H}$ & La/W & reaction & PW91 & PBE      \\
\hline
1H & 2.0   &	3\ce{La2WO6} $\to$ \ce{La6W3O18}	   & 41.42 & 41.42  \\
\hline
\rowcolor{lavender} 3H & 1.8   & (17/2)\ce{La2WO6} + (1/2)\ce{La2W3O12} $\to$ \ce{La18W10O57}  & 64.46 & 77.50 \\
\hline
\rowcolor{lavender} 3H & 1.8   & 8\ce{La2WO6} + \ce{La2W2O9} $\to$ \ce{La18W10O57}  & 71.57 & 69.40 \\
\hline
4H & 1.714 & 11\ce{La2WO6} + \ce{La2W3O12} $\to$ \ce{La24W14O78}  & 88.42 & 108.4 \\
\hline
4H & 1.714 & 10\ce{La2WO6} + 2\ce{La2W2O9} $\to$ \ce{La24W14O78}  & 102.6 & 106.9 \\
\hline
\rowcolor{lavender} 5H & 1.765 & 14\ce{La2WO6} + \ce{La2W3O12} $\to$ \ce{La30W17O96}  & 97.54 & 120.6 \\
\hline
\rowcolor{lavender} 5H & 1.765 & 13\ce{La2WO6} + 2\ce{La2W2O9} $\to$ \ce{La30W17O96}  & 111.8 & 119.2 \\
\hline
6H & 1.8   & 17\ce{La2WO6} + \ce{La2W3O12} $\to$ \ce{La36W20O114} & 114.2 & 140.2 \\
\hline
6H & 1.8   & 16\ce{La2WO6} + 2\ce{La2W2O9} $\to$ \ce{La36W20O114} & 128.4 & 138.8 \\
\hline
\rowcolor{lavender} 7H & 1.75  & (39/2)\ce{La2WO6} + (3/2)\ce{La2W3O12} $\to$ \ce{La42W24O135} & 151.0 & 181.9 \\
\hline
\rowcolor{lavender} 7H & 1.75  & 18\ce{La2WO6} + 3\ce{La2W2O9} $\to$ \ce{La42W24O135} & 172.4 & 179.7 \\
\hline
\end{tabular}}
\caption{\label{t3}\small{DFT-calculated standard enthalpies of~formation, $\Delta_{f}H^{\circ}$, of~the $\mathrm{(N \times H)}$-polytypes at~$T = 0\:\mathrm{K}$, $p = 0\:\mathrm{bar}$ in~standard units of~$\mathrm{kJ/mol}$ ($\mathrm{mol}$ of~a formula unit of~the reaction product; $1\:\mathrm{eV/at.} \approx 96.485\:\mathrm{kJ/mol}$, $1\:\mathrm{kcal/mol} = 4.184\:\mathrm{kJ/mol}$) in the \texttt{PW91} and \texttt{PBE} \texttt{VASP} setups. A positive sign of~$\Delta_{f}H^{\circ}$ stands for~an~endothermic process, i.e. a \emph{metastable} final product. The \ce{La2WO6} reactant on the left-hand side is the stable polymorph \ce{La2WO6}.oP36 from the convex hull.}}
\end{table}

\subsection{\label{unusual}Unusual W Coordination Polyhedra}

\indent The trigonal prism is an unusual \ce{WO6} configuration; to the best of our knowledge it has not been observed in any other tungsten oxide except for \ce{La18W10O57} or \ce{Ce18W10O57}, \cite{5H, la18w10o57, ce18w10o57}, and the class of halotungstates, such as \ce{La3WO6Cl3} or \ce{Pr3WO6Cl3}, \cite{prism-review, parise, dorn, brixner}. While the fundamental stoichiometric \ce{La2WO6} variant of~the structure has neither twinned \ce{W2O9} units nor genuine \ce{WO6} prisms, placement of an extra \ce{WO3} molecule next to~a~\ce{WO6} octahedron -- to form a twinned \ce{W2O9} polyhedron -- causes a spreading distortion of the La sublattice, leading to a deformation of~\ce{WO6} in~the nearby column, which consequently acquires a trigonal-prismatic shape. \\
\indent The emergent \ce{WO6} trigonal prism, mediated by modulation of the La sublattice, will in turn attract other \ce{W2O9} units to~the same (001) layer, resulting in~the twinned \ce{W2O9} units with the same $c$ coordinate in~the laterally-neighbouring unit cells. This is also essential for the manifested $c$ period of the $(\mathrm{N\times H})$-polytypes, since the \ce{W2O9} units created at random in~the lateral cells would destroy the expected superstructures portrayed in \autoref{schematic}, which constitute the main subject of this work. \\

\indent Face-sharing \ce{WO6} octahedra have been reported in various Ba tungstates, such as \ce{Ba3W2O9}, \cite{poeppelmeier}, or \ce{Ba3Fe2WO9}, \cite{ivanov}, or even in some oxygen-deficient \ce{WO3}\textsubscript{-x} with crystallographic shear planes, \cite{wo3-share,kieslich,magneli}. In~addition, face-sharing octahedra have been evidenced in a couple of thorium tungstates: \ce{A6Th6(WO4)14O} for A = K and Rb, \cite{th-tun}. Such face-shared coordination environments of~W are rather rare because of the electrostatic repulsion between the highly charged metal centers, \cite{ce18w10o57}. In order to increase metal-metal distances, tungsten atoms move away from the ideal centers of the octahedra, \cite{la18w10o57}. In~\ce{Ce18W10O57}, a $|W-W| = 2.939$\AA\, distance is expected from diffraction experiments, \cite{ce18w10o57}. In \ce{La18W10O57}, similar W-W separations are considered based on experimental data, ranging from $|W-W| = 2.902\:$\AA\, to~$2.936\:$\AA, \cite{5H}. This is confirmed by~our DFT calculations, wherein~$|W-W| = 2.921\:$\AA\, in 4H, $|W-W| = 2.924-2.938\:$\AA\, in~5H, $|W-W| = 2.940\:$\AA\, in 6H and $|W-W| = 2.924-2.932\:$\AA\, in 7H-polytype. A similar distance was also observed inside~face-shared octahedra of \ce{Ba3W2O9}: $|W-W| = 2.94\:$\AA, \cite{poeppelmeier}. The shortest W-W distance of~$2.70\:$\AA\, has been observed in~\ce{Bi2WO6}, though between~W atoms in octahedra sharing vertices, \cite{bi2wo6}. The present DFT study corroborates previous results of diffraction experiments \cite{5H, la18w10o57, ce18w10o57, voronkova1, voronkova2} regarding the 6H-polytype (either \ce{La18W10O57} or \ce{Ce18W10O57}) as a mixture of two phases: \ce{La3WO6Cl3} for trigonal prisms and \ce{Ba3W2O9} for face-sharing octahedra. \\

\subsection{\label{experiment}Further Comparison with Experiment}

\indent The comparison of experimental lattice parameters, cell volumes and space groups with~our results is given in \autoref{t4}. Except for older works \cite{voronkova1,voronkova2} all sources predict non-centrosymmetric hexagonal or trigonal space groups of higher symmetry than the present DFT calculations. For completeness, also the study \cite{ce18w10o57} states non-centrosymmetric hexagonal space group no.~190 ($P\bar{6}2c$) for 6H-polytype of \ce{Ce18W10O57}. Higher symmetry of experimental data is probably caused by atomic-position averaging over whole samples, which are \emph{always} a mixture of at least two different polytypes, 5H and 6H, \cite{voronkova1,voronkova2}. All attempts to~obtain single crystals (of \ce{La18W10O57}) were unsuccessful, \cite{5H,la18w10o57,ce18w10o57}. \\
\indent In agreement with common knowledge, \cite{dft-lattice}, DFT predicts lattice parameters (bond lengths) systematically larger by $\approx 1\%$ compared with experiment. The electronic band gap of the $\mathrm{(N\times H)}$-polytypes has not been measured and there is not even a general consensus on~the type of their conductivity. \\
\indent Kovalevsky et al., \cite{kovalevsky}, claim mixed oxygen-ionic and electronic conductivity of the related compound La\textsubscript{2}W\textsubscript{1.25}O\textsubscript{6.75} in air ambient at $1170\:\mathrm{K}$. This observation was partially challenged by the study \cite{la18w10o57} of Chambrier et al., stating that electronic conductivity prevails under~the same conditions in \ce{La18W10O57} and that there is practically no ionic conductivity. Remarkably, Chambrier et al. state exactly the same value of the total conductivity measured by the same procedure as Kovalevsky et al., $1.3\times10^{-4}\:\mathrm{S/cm}$, under~the same conditions ($1170\:\mathrm{K}$ in air), but their sample has been prepared by a distinct method and possesses slightly different composition. Unlike Chambrier and Kovalevsky, Voronkova et~al. regard lanthanum oxytungstates as insulators, \cite{voronkova1}. Despite underestimation of the measured gaps by about 40\% within the Kohn-Sham framework, \cite{feliciano-giustino}, our 5H- and 6H-polytypes both exhibit a $2.8\:\mathrm{eV}$-wide band gap. \\
\indent The structure of Kovalevsky's sample has not been determined and only the~PDF-card no.~00-032-0503 produced by Voronkova et al., \cite{voronkova1} was quoted, exhibiting the same XRD pattern. Kovalevsky et al. reported rather large uncertainty of the density of~La\textsubscript{2}W\textsubscript{1.25}O\textsubscript{6.75} measured by the standard pycnometric technique, which comprised only 91\% of the theoretical density calculated from XRD results. Similarly Voronkova et~al. in~their PDF-file state the measured density $7.45\:\mathrm{g/cm^{3}}$, which amounts merely to~93.9\% of~the calculated density $7.93\:\mathrm{g/cm^{3}}$ for the same La\textsubscript{2}W\textsubscript{1.25}O\textsubscript{6.75} composition, whose structure is as yet unknown. \\
\indent Note also that structural models of~the $\mathrm{(N\times H)}$-polytypes suggested by the groups \cite{5H,la18w10o57,ce18w10o57} practically rule out ionic conductivity under~normal conditions since \emph{all} O atoms are bound inside the \ce{WO6} octahedral or trigonal-prismatic units. Moreover, our present structural model rules out Kovalevsky's composition La\textsubscript{2}W\textsubscript{1.25}O\textsubscript{6.75} with 55.55 molar \% \ce{WO3}, since no more \ce{WO3} units than in the 4H-polytype corresponding to 53.85 molar \% \ce{WO3} can be contained within the sample. Our result agrees with the claim of~Chambrier et al., \cite{la18w10o57}, that hexagonal compounds are observed only in the region 52.63--53.3 molar \% \ce{WO3}. The lower boundary, 52.63 molar \% \ce{WO3}, corresponds to the 6H-polytype with~\ce{La18W10O57} stoichiometry. \\
\indent The current mutually contradicting data on the conductivity of the $\mathrm{(N\times H)}$-polytypes could serve as an impetus to the future experimental preparation of ionic conductors from~this family, e.g. via the experimental method suggested by Kovalevsky et al., \cite{kovalevsky}, and subsequent structure determination of the same samples, since their structure and mechanism of conductivity remain unknown.

\begin{table}[h!]
	\centering
	\small{
		\begin{tabular}{|c|c|l|c|c|c|}
			\hline
			\rowcolor{olive9}sample    & source & space group & $a$ [\AA] &  $c$ [\AA] & $V$ [\AA$\mathrm{{}^{3}/cell}$] \\
			\hline
			4H     &  PW91  &  190 ($P\bar{6}2c$) &  9.121     &  22.23     & 1602  \\
			\hline
			4H     &  PBE   &  190 ($P\bar{6}2c$) &  9.127     &  22.25     & 1605  \\
			\hline
			4H     &  \cite{voronkova2} &  194 ($P\bar{6}_{3}/mmc$) &  --    &  22.01     & --  \\
			\hline
			\rowcolor{lavender} 5H     &  PW91  &  143 ($P3$) &  9.111    &  27.60     & 1983  \\
			\hline
			\rowcolor{lavender} 5H     &  PBE   &  143 ($P3$) &  9.116    &  27.62     & 1988  \\
			\hline
			\rowcolor{lavender} 5H     &  \cite{voronkova2} &  191 ($P6/mmm$) &  9.030     & 27.20      & 1921  \\
			\hline
			\rowcolor{lavender} 5H     &  \cite{5H}         &  150 ($P321$) &  9.036     & 27.34      & 1933  \\
			\hline
			6H     &  PW91  &  159 ($P31c$) &  9.105    &  32.95     & 2366  \\
			\hline
			6H     &  PBE   &  159 ($P31c$) &  9.111    &  32.98     & 2372  \\
			\hline
			6H     &  \cite{la18w10o57} &  190 ($P\bar{6}2c$) &  9.045     &  32.68     & 2316  \\
			\hline
			6H     &  \cite{voronkova1} &  194 ($P\bar{6}_{3}/mmc$) &  9.039     &  32.60--33.65 & 2319  \\
			\hline
			6H     &  \cite{5H}         &  190 ($P\bar{6}2c$) &  9.032     &  32.65     & 2307  \\
			\hline
			6H     &  \cite{5H-again}   &  190 ($P\bar{6}2c$) &  9.040     &  32.68     & 2313  \\
			\hline
			\rowcolor{lavender} 7H     &  PW91  &  143 ($P3$) &  9.130    &  38.81     & 2802  \\
			\hline
			\rowcolor{lavender} 7H     &  PBE   &  143 ($P3$) &  9.137    &  38.84     & 2808  \\
			\hline
			\rowcolor{lavender} 7H     &  \cite{voronkova2} &  187 ($P\bar{6}m2$)  &  --        &  38.40     &     --   \\
			\hline
	\end{tabular}}
	\caption{\label{t4}\small{Comparison of the calculated lattice parameters and cell volumes of various $\mathrm{(N\times H)}$-polytypes with \ce{La2WO6}-related chemical compositions in the present DFT study (PW91 vs.~PBE setup) with experimental diffraction data from various groups.}}
\end{table}


\section{\label{conclusions}Conclusions}

\indent We have refined the structure of the experimentally observed 5H-polytype with the composition \ce{La30W17O96} and formulated simple empirical rules, \autoref{ttt}, leading to 3H--7H-polytypes of the \ce{La2WO6} family. By explicit construction of~the ternary La-O-W convex envelope \emph{via} DFT, the energetics of all these compounds at $T=0\:\mathrm{K}$, $p=0\:\mathrm{kbar}$ was assessed, confirming the small distance of all the polytypes from the hull, $\Delta E \approx 9\:\mathrm{meV/at.} \cong 100\:\mathrm{K}$. According to DFT, the ground state of~the family is represented by yet \emph{hypothetical} tungstate \ce{La2WO6}.oP36, obtained by substitution of Gd by La in~\ce{Gd2WO6}, closely related to the basic building H-block of the polytypes. The ground state lies merely $\Delta E \approx 5\:\mathrm{meV/at.}$ below another hypothetical tungstate \ce{La2WO6}.mC72, which has been experimentally prepared as \ce{Sm2MoO6}. This information may be useful for crystallographers and chemists searching for new oxytungstate polymorphs.


\begin{acknowledgments}
We acknowledge the National Grant Scheme Vega 2/0085/22. We thank Dr. Mariana Derzsi for many useful discussions. The research results were obtained using the computational resources procured in the national project the National Competence Centre for High Performance Computing (project code: 311070AKF2) funded by the European Regional Development Fund and the EU Structural Funds Informatization of Society, Operational Program Integrated Infrastructure.
\end{acknowledgments}



\bibliography{Bibliography}

\end{document}


\pagestyle{empty}
\definecolor{ruz}{rgb}{1,0,0.5}
\definecolor{tmod}{RGB}{0,50,150}
\definecolor{tzel}{RGB}{0,200,200}
\definecolor{zel}{RGB}{0,100,50}
\definecolor{fialova}{rgb}{0.5,0,0.5}
\definecolor{white}{rgb}{1.0,1.0,1.0}
\definecolor{tblue}{RGB}{0,0,130}
\definecolor{seda}{RGB}{235,235,235}
\definecolor{sseda}{RGB}{248,248,248}
\definecolor{azur}{RGB}{212,255,230}
\definecolor{pink}{RGB}{242,220,255}
\definecolor{grey}{RGB}{255,240,240}
\definecolor{lavender}{rgb}{0.9, 0.9, 0.98}

\hyphenation{Di-me-ri-sa-ti-on alu-mi-no-si-li-ca-tes}

\numberwithin{equation}{section}
\numberwithin{figure}{section}
\numberwithin{table}{section}












\thispagestyle{empty}
\hypertarget{tocc}{}

\begin{center}
	\noindent {\Huge{\textbf{Family of $\mathrm{\mathbf{(N \times H)}}$-Polytypes}}} \\
\end{center}

\vspace{-0.7cm}

\begin{center}
	\noindent {\Huge{\textbf{with \ce{La2WO6}-Related Stoichiometry}}} \\
\end{center}

\smallskip

\begin{center}
	\noindent {\large{\textbf{Eva Posp\'{i}\v{s}ilov\'{a}, Marek Mihalkovi\v{c}, Na\v{d}a Beronsk\'{a}}}} \\
\end{center}

\vspace{-0.7cm}

\begin{center}
	\noindent {\large{\textbf{and Marek Gebura}}} \\
\end{center}

\smallskip

\begin{center}
\noindent {\Huge{\textbf{Supplemental Material}}} \\
\end{center}

\vspace{1.0cm}

\noindent \underline{Note:} Invisible hyperlinks (white places) at the corners of \emph{every} page of this Supplemental Material (SM) bring you back to this Table of Contents (TOC). \\

\vspace{0.5cm}

\indent 
\noindent \hspace{-0.76cm} {\Large{\textbf{Table of Contents}}} \\
\label{sec:reco}
\vspace{0.5cm}

\noindent \hyperlink{s1}{\large{\textbf{S1. 1H-polytype CONTCAR}}} \\

\noindent \hyperlink{s2}{\large{\textbf{S2. 3H-polytype CONTCAR}}} \\

\noindent \hyperlink{s3}{\large{\textbf{S3. 4H-polytype CONTCAR}}} \\

\noindent \hyperlink{s4}{\large{\textbf{S4. 5H-polytype CONTCAR}}} \\

\noindent \hyperlink{s5}{\large{\textbf{S5. 6H-polytype CONTCAR}}} \\

\noindent \hyperlink{s6}{\large{\textbf{S6. 7H-polytype CONTCAR}}} \\


\newpage
\hypertarget{s1}{}
\noindent {\Large{\textbf{S1. 1H-polytype CONTCAR}}} \\
\label{sec:1h}

\vspace{0.25cm}

\begin{tabular}{lll}
1H-polytype: & La6O18W3  &  \\
1.0  &  &  \\
9.310952  &  -0.001138  &  0.000316  \\
-4.656460  &  8.065098  &  0.003871  \\
0.002547  &  -0.001611  &  5.462956  \\
La  &  O  &  W   \\
6  &  18  &  3   \\
Direct  &  &  \\
0.926896  &  0.594926  &  0.464658  \\
0.667931  &  0.075618  &  0.465101  \\
0.407966  &  0.334566  &  0.465281  \\
0.666457  &  0.731979  &  0.963751  \\
0.270028  &  0.936747  &  0.963808  \\
0.066148  &  0.336467  &  0.963935  \\
0.898950  &  0.807696  &  0.252217  \\
0.195352  &  0.093855  &  0.252655  \\
0.909185  &  0.104306  &  0.252888  \\
0.525761  &  0.802869  &  0.260452  \\
0.277978  &  0.477064  &  0.260204  \\
0.199770  &  0.724662  &  0.259645  \\
0.812169  &  0.301414  &  0.744028  \\
0.700502  &  0.512530  &  0.743363  \\
0.489437  &  0.190014  &  0.744000  \\
0.056819  &  0.866336  &  0.670622  \\
0.136103  &  0.192680  &  0.671281  \\
0.809623  &  0.945559  &  0.670796  \\
0.432720  &  0.573061  &  0.680067  \\
0.429081  &  0.862049  &  0.679507  \\
0.140145  &  0.569419  &  0.679353  \\
0.523947  &  0.156257  &  0.183865  \\
0.632546  &  0.478638  &  0.183504  \\
0.846297  &  0.370171  &  0.183683  \\
0.000964  &  0.001836  &  0.459342  \\
0.334276  &  0.668248  &  0.470997  \\
0.667513  &  0.334818  &  0.964284  \\
\end{tabular}


\newpage
\hypertarget{s2}{}
\noindent {\Large{\textbf{S2. 3H-polytype CONTCAR}}} \\
\label{sec:3h}

\vspace{0.25cm}

\begin{tabular}{lll}
3H-polytype: & La18O57W10  &  \\
1.0  &  &  \\
9.151583  &  0.000034  &  0.000032  \\
-4.575762  &  7.925595  &  -0.000006  \\
0.000075  &  -0.000007  &  16.595856  \\
La  &  O  &  W   \\
18  &  57  &  10   \\
Direct  &  &  \\
0.009344  &  0.259289  &  0.095260  \\
0.740697  &  0.750060  &  0.095264  \\
0.249922  &  0.990626  &  0.095260  \\
0.429903  &  0.372487  &  0.257599  \\
0.942585  &  0.570114  &  0.257598  \\
0.627516  &  0.057422  &  0.257595  \\
0.243210  &  0.965233  &  0.427717  \\
0.034784  &  0.277991  &  0.427720  \\
0.722024  &  0.756802  &  0.427721  \\
0.429939  &  0.370932  &  0.598085  \\
0.629071  &  0.059014  &  0.598084  \\
0.940979  &  0.570070  &  0.598086  \\
0.012409  &  0.260351  &  0.759233  \\
0.739639  &  0.752091  &  0.759233  \\
0.247916  &  0.987583  &  0.759236  \\
0.393391  &  0.443656  &  0.927496  \\
0.050247  &  0.606605  &  0.927498  \\
0.556318  &  0.949723  &  0.927496  \\
0.520229  &  0.718255  &  0.010976  \\
0.198024  &  0.479760  &  0.010971  \\
0.281740  &  0.801965  &  0.010969  \\
0.858818  &  0.392540  &  0.027478  \\
0.533706  &  0.141169  &  0.027473  \\
0.607435  &  0.466273  &  0.027481  \\
0.405747  &  0.537265  &  0.145994  \\
0.131493  &  0.594234  &  0.145989  \\
0.462695  &  0.868462  &  0.146000  \\
0.800457  &  0.283188  &  0.171956  \\
0.482700  &  0.199525  &  0.171950  \\
0.716805  &  0.517281  &  0.171956  \\
0.158356  &  0.168540  &  0.188437  \\
0.010155  &  0.841648  &  0.188432  \\
0.831469  &  0.989841  &  0.188431  \\
0.173357  &  0.467799  &  0.295600  \\
0.294445  &  0.826660  &  0.295599  \\
0.532231  &  0.705566  &  0.295591  \\
0.177427  &  0.148590  &  0.337868  \\
\end{tabular}

\begin{tabular}{lll}
0.851411  &  0.028852  &  0.337866  \\
0.971160  &  0.822588  &  0.337866  \\
0.823269  &  0.320676  &  0.352345  \\
0.497413  &  0.176756  &  0.352341  \\
0.679348  &  0.502609  &  0.352345  \\
0.361001  &  0.520032  &  0.428973  \\
0.159058  &  0.639034  &  0.428975  \\
0.479996  &  0.840967  &  0.428969  \\
0.822160  &  0.318748  &  0.502997  \\
0.496594  &  0.177859  &  0.502996  \\
0.681276  &  0.503425  &  0.502999  \\
0.176673  &  0.149714  &  0.516881  \\
0.973043  &  0.823345  &  0.516880  \\
0.850302  &  0.026974  &  0.516877  \\
0.171313  &  0.468556  &  0.562314  \\
0.531457  &  0.702759  &  0.562305  \\
0.297260  &  0.828708  &  0.562308  \\
0.160330  &  0.165520  &  0.666549  \\
0.834486  &  0.994840  &  0.666545  \\
0.005175  &  0.839677  &  0.666547  \\
0.484175  &  0.196169  &  0.682963  \\
0.711970  &  0.515832  &  0.682960  \\
0.803822  &  0.288026  &  0.682959  \\
0.390480  &  0.524821  &  0.711226  \\
0.134327  &  0.609515  &  0.711223  \\
0.475175  &  0.865662  &  0.711229  \\
0.531751  &  0.141994  &  0.827998  \\
0.610227  &  0.468260  &  0.827999  \\
0.858011  &  0.389787  &  0.827993  \\
0.175335  &  0.152278  &  0.851919  \\
0.976907  &  0.824658  &  0.851915  \\
0.847719  &  0.023080  &  0.851914  \\
0.519394  &  0.714292  &  0.845779  \\
0.194874  &  0.480589  &  0.845775  \\
0.285679  &  0.805113  &  0.845774  \\
0.175155  &  0.152594  &  0.002056  \\
0.847385  &  0.022557  &  0.002052  \\
0.977429  &  0.824829  &  0.002054  \\
0.333319  &  0.666651  &  0.091422  \\
0.666650  &  0.333324  &  0.088392  \\
0.999996  &  0.000013  &  0.263796  \\
0.333359  &  0.666685  &  0.340740  \\
0.666675  &  0.333345  &  0.427675  \\
0.333335  &  0.666670  &  0.517378  \\
0.000002  &  0.000012  &  0.590804  \\
0.666655  &  0.333343  &  0.767153  \\
0.333319  &  0.666661  &  0.765195  \\
0.999990  &  0.000001  &  0.927134  \\	
\end{tabular}


\newpage
\hypertarget{s3}{}
\noindent {\Large{\textbf{S3. 4H-polytype CONTCAR}}} \\
\label{sec:4h}

\vspace{0.25cm}

\begin{tabular}{lll}
4H-polytype: & La24O78W14  &  \\
1.0  &  &  \\
9.127376  &  0.000016  &  0.000016  \\
-4.563674  &  7.904527  &  0.000002  \\
0.000050  &  0.000009  &  22.251872  \\
La  &  O  &  W   \\
24  &  78  &  14   \\
Direct  &  &  \\
0.255052  &  0.998428  &  0.004021  \\
0.743382  &  0.744923  &  0.004023  \\
0.001570  &  0.256610  &  0.004019  \\
0.570135  &  0.943925  &  0.127839  \\
0.373796  &  0.429866  &  0.127845  \\
0.056067  &  0.626201  &  0.127843  \\
0.757127  &  0.723022  &  0.254359  \\
0.965901  &  0.242911  &  0.254362  \\
0.277025  &  0.034136  &  0.254362  \\
0.570299  &  0.943119  &  0.380523  \\
0.056942  &  0.627222  &  0.380530  \\
0.372828  &  0.429754  &  0.380530  \\
0.998585  &  0.255165  &  0.504040  \\
0.744869  &  0.743429  &  0.504038  \\
0.256568  &  0.001437  &  0.504040  \\
0.429833  &  0.373765  &  0.627852  \\
0.943917  &  0.570147  &  0.627854  \\
0.626216  &  0.056049  &  0.627855  \\
0.242874  &  0.965870  &  0.754373  \\
0.034078  &  0.276950  &  0.754379  \\
0.722990  &  0.757079  &  0.754369  \\
0.429709  &  0.372720  &  0.880533  \\
0.627254  &  0.056964  &  0.880534  \\
0.943010  &  0.570246  &  0.880535  \\
0.866731  &  0.454389  &  0.043703  \\
0.545613  &  0.412358  &  0.043705  \\
0.587626  &  0.133253  &  0.043709  \\
0.190327  &  0.479658  &  0.068061  \\
0.520326  &  0.710652  &  0.068060  \\
0.289336  &  0.809670  &  0.068060  \\
0.174535  &  0.153348  &  0.073543  \\
0.978823  &  0.825469  &  0.073543  \\
0.846655  &  0.021175  &  0.073538  \\
0.825929  &  0.292976  &  0.155766  \\
0.707056  &  0.532949  &  0.155763  \\
0.467066  &  0.174091  &  0.155766  \\
0.147574  &  0.177684  &  0.185488  \\
\end{tabular}

\begin{tabular}{lll}
0.822318  &  0.969893  &  0.185483  \\
0.030116  &  0.852436  &  0.185486  \\	
0.176370  &  0.496564  &  0.198391  \\
0.503454  &  0.679812  &  0.198388  \\
0.320200  &  0.823653  &  0.198387  \\
0.842513  &  0.479968  &  0.255018  \\
0.637477  &  0.157543  &  0.255024  \\
0.520085  &  0.362573  &  0.255018  \\
0.504050  &  0.681481  &  0.310247  \\
0.177430  &  0.496007  &  0.310249  \\
0.318584  &  0.822623  &  0.310247  \\
0.823014  &  0.971426  &  0.322665  \\
0.028639  &  0.851634  &  0.322663  \\
0.148427  &  0.177058  &  0.322668  \\
0.826858  &  0.294139  &  0.354191  \\
0.467297  &  0.173239  &  0.354186  \\
0.705942  &  0.532801  &  0.354188  \\
0.981953  &  0.827237  &  0.434708  \\
0.172836  &  0.154761  &  0.434708  \\
0.845308  &  0.018123  &  0.434704  \\
0.520299  &  0.710807  &  0.441001  \\
0.190500  &  0.479735  &  0.440994  \\
0.289246  &  0.809530  &  0.440998  \\
0.865918  &  0.462103  &  0.465742  \\
0.596183  &  0.134096  &  0.465737  \\
0.537926  &  0.403849  &  0.465742  \\
0.412155  &  0.545433  &  0.543679  \\
0.133276  &  0.587858  &  0.543677  \\
0.454580  &  0.866749  &  0.543685  \\
0.809628  &  0.289277  &  0.568052  \\
0.479641  &  0.190368  &  0.568049  \\
0.710717  &  0.520361  &  0.568052  \\
0.153398  &  0.174479  &  0.573562  \\
0.021081  &  0.846593  &  0.573564  \\
0.825510  &  0.978912  &  0.573562  \\
0.174033  &  0.467037  &  0.655778  \\
0.292952  &  0.825872  &  0.655781  \\
0.532875  &  0.706962  &  0.655780  \\
0.177665  &  0.147545  &  0.685503  \\
0.852415  &  0.030089  &  0.685503  \\
0.969858  &  0.822296  &  0.685501  \\
0.823651  &  0.320209  &  0.698397  \\
0.496553  &  0.176308  &  0.698387  \\
0.679745  &  0.503393  &  0.698390  \\
0.362496  &  0.520015  &  0.755052  \\
0.157511  &  0.637454  &  0.755058  \\
0.479930  &  0.842438  &  0.755051  \\
0.822448  &  0.318336  &  0.810260  \\
0.495880  &  0.177479  &  0.810252  \\
\end{tabular}

\begin{tabular}{lll}
0.681591  &  0.504053  &  0.810252  \\
0.176949  &  0.148414  &  0.822654  \\
0.971456  &  0.822982  &  0.822663  \\
0.851547  &  0.028498  &  0.822656  \\
0.173070  &  0.467291  &  0.854228  \\
0.532702  &  0.705746  &  0.854222  \\
0.294240  &  0.826913  &  0.854226  \\
0.154800  &  0.172676  &  0.934699  \\
0.827292  &  0.982085  &  0.934700  \\
0.017885  &  0.845171  &  0.934708  \\
0.479742  &  0.190404  &  0.941009  \\
0.809580  &  0.289308  &  0.941014  \\
0.710656  &  0.520238  &  0.941005  \\
0.134165  &  0.596705  &  0.965760  \\
0.403265  &  0.537434  &  0.965758  \\
0.462547  &  0.865818  &  0.965760  \\
0.333327  &  0.666653  &  0.006814  \\
0.000004  &  0.000002  &  0.129989  \\
0.666672  &  0.333336  &  0.189395  \\
0.333350  &  0.666694  &  0.254133  \\
0.666710  &  0.333399  &  0.320674  \\
0.000033  &  0.000043  &  0.378333  \\
0.333350  &  0.666693  &  0.502263  \\
0.666669  &  0.333338  &  0.506810  \\
0.999990  &  0.999988  &  0.629989  \\
0.333282  &  0.666616  &  0.689427  \\
0.666643  &  0.333293  &  0.754135  \\
0.333339  &  0.666650  &  0.820717  \\
0.999986  &  0.999963  &  0.878322  \\
0.666655  &  0.333322  &  0.002288  \\	
\end{tabular}


\newpage
\hypertarget{s4}{}
\noindent {\Large{\textbf{S4. 5H-polytype CONTCAR}}} \\
\label{sec:5h}

\vspace{0.25cm}

\begin{tabular}{lll}
5H-polytype: & La30O96W17  &  \\
1.0  &  &  \\
9.117117  &  0.000011  &  -0.000005  \\
-4.558549  &  7.895637  &  0.000009  \\
-0.000010  &  0.000011  &  27.621147  \\
La  &  O  &  W   \\
30  &  96  &  17   \\
Direct  &  &  \\
0.266672  &  0.014406  &  0.003313  \\
0.747729  &  0.733327  &  0.003312  \\
0.985593  &  0.252265  &  0.003313  \\
0.569454  &  0.946421  &  0.099736  \\
0.376965  &  0.430545  &  0.099736  \\
0.053579  &  0.623032  &  0.099736  \\
0.757384  &  0.722934  &  0.203074  \\
0.965557  &  0.242618  &  0.203072  \\
0.277058  &  0.034439  &  0.203075  \\
0.569153  &  0.942337  &  0.304488  \\
0.057657  &  0.626815  &  0.304488  \\
0.373187  &  0.430842  &  0.304489  \\
0.995593  &  0.254580  &  0.403351  \\
0.745424  &  0.741008  &  0.403351  \\
0.258987  &  0.004412  &  0.403350  \\
0.430443  &  0.376569  &  0.503383  \\
0.946127  &  0.569557  &  0.503384  \\
0.623429  &  0.053873  &  0.503384  \\
0.243412  &  0.967597  &  0.605889  \\
0.032399  &  0.275815  &  0.605892  \\
0.724180  &  0.756593  &  0.605891  \\
0.428590  &  0.369415  &  0.708528  \\
0.630595  &  0.059183  &  0.708528  \\
0.940816  &  0.571407  &  0.708529  \\
0.036432  &  0.274898  &  0.803290  \\
0.725094  &  0.761541  &  0.803290  \\
0.238465  &  0.963564  &  0.803290  \\
0.418859  &  0.398594  &  0.899973  \\
0.979729  &  0.581138  &  0.899973  \\
0.601405  &  0.020269  &  0.899973  \\
0.867335  &  0.448220  &  0.031949  \\
0.551771  &  0.419109  &  0.031947  \\
0.580881  &  0.132658  &  0.031948  \\
0.195762  &  0.480691  &  0.051426  \\
0.519305  &  0.715071  &  0.051426  \\
0.284924  &  0.804234  &  0.051426  \\
0.171039  &  0.157771  &  0.059147  \\
\end{tabular}

\begin{tabular}{lll}
0.986728  &  0.828956  &  0.059147  \\
0.842224  &  0.013267  &  0.059147  \\
0.823369  &  0.289330  &  0.123726  \\
0.710651  &  0.534047  &  0.123726  \\
0.465942  &  0.176615  &  0.123725  \\
0.148508  &  0.177821  &  0.149071  \\
0.822184  &  0.970690  &  0.149072  \\
0.029309  &  0.851492  &  0.149071  \\
0.174694  &  0.497267  &  0.157812  \\
0.502730  &  0.677420  &  0.157813  \\
0.322577  &  0.825304  &  0.157813  \\
0.843581  &  0.477540  &  0.204027  \\
0.633959  &  0.156421  &  0.204028  \\
0.522463  &  0.366040  &  0.204029  \\
0.504450  &  0.683027  &  0.248138  \\
0.178582  &  0.495549  &  0.248137  \\
0.316969  &  0.821416  &  0.248137  \\
0.822793  &  0.971537  &  0.258120  \\
0.028458  &  0.851268  &  0.258123  \\
0.148732  &  0.177198  &  0.258121  \\
0.825654  &  0.291970  &  0.283974  \\
0.466332  &  0.174348  &  0.283974  \\
0.708031  &  0.533669  &  0.283973  \\
0.984405  &  0.828187  &  0.348173  \\
0.171810  &  0.156230  &  0.348174  \\
0.843765  &  0.015592  &  0.348174  \\
0.519794  &  0.712703  &  0.354244  \\
0.192911  &  0.480211  &  0.354242  \\
0.287298  &  0.807089  &  0.354243  \\
0.866785  &  0.460533  &  0.373364  \\
0.593760  &  0.133214  &  0.373364  \\
0.539464  &  0.406237  &  0.373364  \\
0.424921  &  0.557469  &  0.437892  \\
0.132552  &  0.575086  &  0.437892  \\
0.442538  &  0.867455  &  0.437892  \\
0.807222  &  0.286334  &  0.455997  \\
0.479117  &  0.192777  &  0.455996  \\
0.713663  &  0.520880  &  0.455996  \\
0.151592  &  0.177328  &  0.459850  \\
0.025746  &  0.848416  &  0.459848  \\
0.822670  &  0.974259  &  0.459850  \\
0.177615  &  0.467681  &  0.530880  \\
0.290060  &  0.822382  &  0.530879  \\
0.532316  &  0.709931  &  0.530880  \\
0.179783  &  0.145710  &  0.549492  \\
0.854290  &  0.034079  &  0.549492  \\
0.965920  &  0.820216  &  0.549492  \\
0.826831  &  0.326445  &  0.561524  \\
0.499609  &  0.173168  &  0.561524  \\
\end{tabular}

\begin{tabular}{lll}
0.673553  &  0.500400  &  0.561524  \\
0.362008  &  0.520233  &  0.612291  \\
0.158217  &  0.637991  &  0.612290  \\
0.479765  &  0.841782  &  0.612290  \\
0.819536  &  0.312779  &  0.651732  \\
0.493234  &  0.180466  &  0.651731  \\
0.687218  &  0.506768  &  0.651732  \\
0.173694  &  0.153634  &  0.657905  \\
0.979958  &  0.826311  &  0.657904  \\
0.846377  &  0.020052  &  0.657905  \\
0.170166  &  0.469130  &  0.693179  \\
0.530864  &  0.701042  &  0.693180  \\
0.298960  &  0.829833  &  0.693180  \\
0.169165  &  0.155643  &  0.748017  \\
0.844364  &  0.013510  &  0.748015  \\
0.986494  &  0.830838  &  0.748017  \\
0.490645  &  0.185412  &  0.756618  \\
0.694762  &  0.509360  &  0.756619  \\
0.814594  &  0.305238  &  0.756620  \\
0.368341  &  0.507146  &  0.787075  \\
0.138789  &  0.631652  &  0.787075  \\
0.492851  &  0.861210  &  0.787075  \\
0.500364  &  0.168967  &  0.846050  \\
0.668590  &  0.499634  &  0.846050  \\
0.831031  &  0.331405  &  0.846051  \\
0.179685  &  0.146095  &  0.856947  \\
0.966407  &  0.820315  &  0.856947  \\
0.853907  &  0.033592  &  0.856947  \\
0.524256  &  0.710396  &  0.866574  \\
0.186128  &  0.475741  &  0.866575  \\
0.289602  &  0.813868  &  0.866575  \\
0.143204  &  0.184706  &  0.945685  \\
0.815299  &  0.958496  &  0.945684  \\
0.041505  &  0.856798  &  0.945685  \\
0.472259  &  0.206872  &  0.949861  \\
0.793120  &  0.265382  &  0.949863  \\
0.734612  &  0.527737  &  0.949863  \\
0.133707  &  0.556874  &  0.967241  \\
0.443121  &  0.576836  &  0.967241  \\
0.423160  &  0.866288  &  0.967241  \\
0.333331  &  0.666665  &  0.002180  \\
1.000000  &  0.000001  &  0.104213  \\
0.666663  &  0.333337  &  0.151075  \\
0.333332  &  0.666663  &  0.202309  \\
0.666674  &  0.333325  &  0.256937  \\
0.000010  &  0.999998  &  0.302660  \\
0.333339  &  0.666670  &  0.404182  \\
\end{tabular}

\begin{tabular}{lll}
0.666668  &  0.333325  &  0.406009  \\
0.999998  &  0.000001  &  0.504953  \\
0.333329  &  0.666665  &  0.559277  \\
0.666660  &  0.333340  &  0.606296  \\
0.333327  &  0.666670  &  0.665643  \\
0.000006  &  0.999999  &  0.702829  \\
0.666662  &  0.333337  &  0.805068  \\
0.333322  &  0.666665  &  0.823781  \\
0.999997  &  0.999999  &  0.902673  \\
0.666663  &  0.333330  &  0.995508  \\	
\end{tabular}


\newpage
\hypertarget{s5}{}
\noindent {\Large{\textbf{S5. 6H-polytype CONTCAR}}} \\
\label{sec:6h}

\vspace{0.25cm}

\begin{tabular}{lll}
6H-polytype: & La36O114W20  &  \\
1.0  &  &  \\
9.111240  &  0.000009  &  0.000009  \\
-4.555612  &  7.890583  &  -0.000018  \\
0.000034  &  -0.000055  &  32.983320  \\
La  &  O  &  W   \\
36  &  114  &  20   \\
Direct  &  &  \\
0.988428  &  0.253044  &  0.001468  \\
0.253061  &  0.988458  &  0.501467  \\
0.746940  &  0.735388  &  0.001467  \\
0.011553  &  0.264613  &  0.501465  \\
0.264604  &  0.011532  &  0.001467  \\
0.735393  &  0.746943  &  0.501466  \\
0.756765  &  0.724440  &  0.169476  \\
0.963309  &  0.238814  &  0.334376  \\
0.275566  &  0.032311  &  0.169479  \\
0.761196  &  0.724494  &  0.334374  \\
0.967691  &  0.243236  &  0.169477  \\
0.275515  &  0.036696  &  0.334376  \\
0.036693  &  0.275519  &  0.834374  \\
0.238807  &  0.963308  &  0.834376  \\
0.724483  &  0.761196  &  0.834375  \\
0.724437  &  0.756763  &  0.669478  \\
0.032316  &  0.275562  &  0.669477  \\
0.243240  &  0.967688  &  0.669476  \\
0.571065  &  0.939637  &  0.254653  \\
0.368585  &  0.428944  &  0.254654  \\
0.060384  &  0.631429  &  0.254654  \\
0.631427  &  0.060380  &  0.754651  \\
0.428943  &  0.368578  &  0.754651  \\
0.939618  &  0.571056  &  0.754651  \\
0.379241  &  0.430290  &  0.082338  \\
0.021667  &  0.602642  &  0.415284  \\
0.569691  &  0.948924  &  0.082335  \\
0.397361  &  0.419016  &  0.415285  \\
0.051058  &  0.620746  &  0.082339  \\
0.580993  &  0.978355  &  0.415284  \\
0.978340  &  0.580990  &  0.915284  \\
0.602624  &  0.021650  &  0.915286  \\
0.419003  &  0.397358  &  0.915281  \\
0.430294  &  0.379243  &  0.582334  \\
0.948947  &  0.569710  &  0.582335  \\
0.620763  &  0.051064  &  0.582336  \\
0.826323  &  0.979995  &  0.213327  \\
\end{tabular}

\begin{tabular}{lll}
0.155688  &  0.169694  &  0.288649  \\
0.020012  &  0.846314  &  0.213328  \\
0.830307  &  0.985985  &  0.288649  \\
0.153688  &  0.173679  &  0.213328  \\
0.014019  &  0.844319  &  0.288649  \\
0.844319  &  0.014017  &  0.788649  \\
0.169696  &  0.155689  &  0.788648  \\
0.985983  &  0.830304  &  0.788649  \\
0.979998  &  0.826319  &  0.713327  \\
0.846321  &  0.020008  &  0.713328  \\
0.173685  &  0.153685  &  0.713327  \\
0.499923  &  0.671331  &  0.131867  \\
0.170402  &  0.499366  &  0.370595  \\
0.328657  &  0.828567  &  0.131868  \\
0.500641  &  0.671017  &  0.370594  \\
0.171426  &  0.500065  &  0.131867  \\
0.328980  &  0.829607  &  0.370597  \\
0.829607  &  0.328983  &  0.870595  \\
0.499364  &  0.170398  &  0.870594  \\
0.671012  &  0.500637  &  0.870594  \\
0.671303  &  0.499911  &  0.631866  \\
0.828591  &  0.328689  &  0.631864  \\
0.500087  &  0.171402  &  0.631865  \\
0.844121  &  0.017544  &  0.048510  \\
0.183687  &  0.144818  &  0.453539  \\
0.982449  &  0.826557  &  0.048508  \\
0.855186  &  0.038870  &  0.453538  \\
0.173428  &  0.155861  &  0.048507  \\
0.961133  &  0.816310  &  0.453538  \\
0.816288  &  0.961121  &  0.953539  \\
0.144796  &  0.183687  &  0.953537  \\
0.038851  &  0.855183  &  0.953540  \\
0.017557  &  0.844129  &  0.548508  \\
0.826564  &  0.982445  &  0.548508  \\
0.155871  &  0.173433  &  0.548507  \\
0.197015  &  0.480020  &  0.042366  \\
0.263914  &  0.791253  &  0.458061  \\
0.519948  &  0.716959  &  0.042367  \\
0.208757  &  0.472669  &  0.458061  \\
0.283015  &  0.802955  &  0.042368  \\
0.527333  &  0.736093  &  0.458060  \\
0.736091  &  0.527330  &  0.958062  \\
0.791230  &  0.263898  &  0.958061  \\
0.472661  &  0.208753  &  0.958060  \\
0.480046  &  0.197028  &  0.542366  \\
0.716981  &  0.519961  &  0.542366  \\
0.802982  &  0.283027  &  0.542365  \\
0.820456  &  0.967796  &  0.123483  \\
\end{tabular}

\begin{tabular}{lll}
0.146233  &  0.179682  &  0.379270  \\
0.032199  &  0.852638  &  0.123484  \\
0.820322  &  0.966548  &  0.379269  \\
0.147361  &  0.179537  &  0.123482  \\
0.033459  &  0.853770  &  0.379270  \\
0.853767  &  0.033468  &  0.879271  \\
0.179671  &  0.146225  &  0.879269  \\
0.966543  &  0.820321  &  0.879269  \\
0.967782  &  0.820449  &  0.623483  \\
0.852651  &  0.032212  &  0.623483  \\
0.179550  &  0.147347  &  0.623482  \\
0.506826  &  0.688297  &  0.207518  \\
0.185085  &  0.490949  &  0.295566  \\
0.311718  &  0.818517  &  0.207519  \\
0.509064  &  0.694134  &  0.295565  \\
0.181493  &  0.493183  &  0.207519  \\
0.305875  &  0.814928  &  0.295568  \\
0.814918  &  0.305871  &  0.795565  \\
0.490944  &  0.185081  &  0.795564  \\
0.694129  &  0.509064  &  0.795565  \\
0.688306  &  0.506822  &  0.707516  \\
0.818511  &  0.311698  &  0.707514  \\
0.493182  &  0.181493  &  0.707515  \\
0.467763  &  0.178340  &  0.106494  \\
0.711065  &  0.524761  &  0.388090  \\
0.821655  &  0.289406  &  0.106490  \\
0.475244  &  0.186301  &  0.388089  \\
0.710584  &  0.532222  &  0.106493  \\
0.813705  &  0.288938  &  0.388091  \\
0.288937  &  0.813710  &  0.888094  \\
0.524755  &  0.711060  &  0.888091  \\
0.186297  &  0.475243  &  0.888094  \\
0.178360  &  0.467769  &  0.606491  \\
0.289397  &  0.821644  &  0.606492  \\
0.532239  &  0.710602  &  0.606492  \\
0.133073  &  0.559119  &  0.972016  \\
0.441802  &  0.867494  &  0.527172  \\
0.440864  &  0.573940  &  0.972016  \\
0.132513  &  0.574316  &  0.527172  \\
0.426036  &  0.866894  &  0.972018  \\
0.425693  &  0.558209  &  0.527171  \\
0.558200  &  0.425688  &  0.027172  \\
0.867482  &  0.441787  &  0.027172  \\
0.574298  &  0.132507  &  0.027173  \\
0.559140  &  0.133089  &  0.472017  \\
0.573950  &  0.440867  &  0.472017  \\
0.866917  &  0.426053  &  0.472015  \\
0.829982  &  0.299088  &  0.242793  \\
\end{tabular}

\begin{tabular}{lll}
0.469135  &  0.170054  &  0.242795  \\
0.700951  &  0.530906  &  0.242795  \\
0.530879  &  0.700931  &  0.742793  \\
0.170028  &  0.469114  &  0.742793  \\
0.299067  &  0.829969  &  0.742795  \\
0.637168  &  0.158025  &  0.175003  \\
0.507908  &  0.369611  &  0.321648  \\
0.841989  &  0.479136  &  0.175001  \\
0.630398  &  0.138292  &  0.321650  \\
0.520877  &  0.362848  &  0.175005  \\
0.861719  &  0.492101  &  0.321651  \\
0.492091  &  0.861714  &  0.821654  \\
0.369599  &  0.507905  &  0.821651  \\
0.138277  &  0.630386  &  0.821654  \\
0.158021  &  0.637168  &  0.675003  \\
0.479133  &  0.841982  &  0.675003  \\
0.362835  &  0.520869  &  0.675002  \\
0.000004  &  0.999988  &  0.086231  \\
0.000003  &  0.999998  &  0.417383  \\
0.999982  &  0.000002  &  0.917384  \\
0.999994  &  0.000002  &  0.586230  \\
0.000001  &  0.000006  &  0.250855  \\
0.999999  &  0.000003  &  0.750857  \\
0.333319  &  0.666639  &  0.001055  \\
0.333337  &  0.666678  &  0.496354  \\
0.666657  &  0.333328  &  0.996354  \\
0.666672  &  0.333339  &  0.501055  \\
0.333341  &  0.666659  &  0.168821  \\
0.333344  &  0.666668  &  0.335990  \\
0.666663  &  0.333340  &  0.835989  \\
0.666663  &  0.333336  &  0.668816  \\
0.666671  &  0.333319  &  0.130525  \\
0.666674  &  0.333337  &  0.352269  \\
0.333324  &  0.666671  &  0.852272  \\
0.333335  &  0.666673  &  0.630524  \\
0.666686  &  0.333352  &  0.219646  \\
0.333321  &  0.666673  &  0.719647  \\	
\end{tabular}


\newpage
\hypertarget{s6}{}
\noindent {\Large{\textbf{S6. 7H-polytype CONTCAR}}} \\
\label{sec:7h}

\vspace{0.25cm}

\begin{tabular}{lll}
7H-polytype: & La42O135W24  &  \\
1.0  &  &  \\
9.136872  &  -0.000017  &  -0.000012  \\
-4.568443  &  7.912760  &  0.000023  \\
-0.000051  &  0.000082  &  38.843760  \\
La  &  O  &  W   \\
42  &  135  &  24   \\
Direct  &  &  \\
0.255694  &  0.999584  &  0.503306  \\
0.743885  &  0.744317  &  0.503305  \\
0.000433  &  0.256121  &  0.503306  \\
0.570236  &  0.943478  &  0.573914  \\
0.373242  &  0.429782  &  0.573915  \\
0.056541  &  0.626759  &  0.573913  \\
0.756704  &  0.722653  &  0.646353  \\
0.965950  &  0.243301  &  0.646351  \\
0.277366  &  0.034069  &  0.646351  \\
0.570401  &  0.942131  &  0.718690  \\
0.057825  &  0.628232  &  0.718691  \\
0.371750  &  0.429564  &  0.718691  \\
0.995715  &  0.253641  &  0.788885  \\
0.746320  &  0.742032  &  0.788886  \\
0.257920  &  0.004233  &  0.788883  \\
0.429998  &  0.374558  &  0.859859  \\
0.944572  &  0.569975  &  0.859861  \\
0.625392  &  0.055383  &  0.859861  \\
0.243532  &  0.966020  &  0.932423  \\
0.033964  &  0.277519  &  0.932422  \\
0.722464  &  0.756460  &  0.932426  \\
0.429724  &  0.372963  &  0.432700  \\
0.627060  &  0.056753  &  0.432702  \\
0.943246  &  0.570278  &  0.432702  \\
0.009230  &  0.258993  &  0.218076  \\
0.741026  &  0.750269  &  0.218074  \\
0.249763  &  0.990778  &  0.218070  \\
0.430110  &  0.372993  &  0.287671  \\
0.942892  &  0.569913  &  0.287666  \\
0.627031  &  0.057116  &  0.287666  \\
0.242871  &  0.965359  &  0.360467  \\
0.034669  &  0.277496  &  0.360467  \\
0.722515  &  0.757144  &  0.360468  \\
0.429949  &  0.370018  &  0.005388  \\
0.629990  &  0.059979  &  0.005385  \\
0.939998  &  0.570069  &  0.005386  \\
\end{tabular}

\begin{tabular}{lll}
0.015655  &  0.261622  &  0.074032  \\
0.738397  &  0.754085  &  0.074034  \\
0.245917  &  0.984371  &  0.074034  \\
0.393937  &  0.442778  &  0.146296  \\
0.048813  &  0.606095  &  0.146302  \\
0.557246  &  0.951217  &  0.146302  \\
0.866286  &  0.457401  &  0.525747  \\
0.542598  &  0.408884  &  0.525747  \\
0.591120  &  0.133713  &  0.525747  \\
0.190331  &  0.479919  &  0.539635  \\
0.520082  &  0.710414  &  0.539636  \\
0.289585  &  0.809655  &  0.539630  \\
0.173843  &  0.153654  &  0.542933  \\
0.979824  &  0.826170  &  0.542933  \\
0.846351  &  0.020186  &  0.542933  \\
0.826199  &  0.293703  &  0.590002  \\
0.706325  &  0.532506  &  0.590005  \\
0.467530  &  0.173840  &  0.590005  \\
0.147328  &  0.177819  &  0.607001  \\
0.822186  &  0.969509  &  0.607001  \\
0.030508  &  0.852689  &  0.607002  \\
0.176349  &  0.496858  &  0.614335  \\
0.503155  &  0.679507  &  0.614336  \\
0.320508  &  0.823646  &  0.614335  \\
0.841747  &  0.480529  &  0.646931  \\
0.638787  &  0.158248  &  0.646932  \\
0.519473  &  0.361214  &  0.646932  \\
0.503972  &  0.681787  &  0.678440  \\
0.177819  &  0.496030  &  0.678438  \\
0.318216  &  0.822192  &  0.678438  \\
0.823109  &  0.971746  &  0.685164  \\
0.028248  &  0.851358  &  0.685164  \\
0.148633  &  0.176874  &  0.685164  \\
0.827895  &  0.295841  &  0.703784  \\
0.467943  &  0.172113  &  0.703784  \\
0.704170  &  0.532026  &  0.703787  \\
0.984521  &  0.828817  &  0.749315  \\
0.171154  &  0.155671  &  0.749315  \\
0.844293  &  0.015450  &  0.749315  \\
0.519750  &  0.709919  &  0.753445  \\
0.190182  &  0.480198  &  0.753444  \\
0.290021  &  0.809769  &  0.753448  \\
0.864518  &  0.468295  &  0.767607  \\
0.603758  &  0.135384  &  0.767608  \\
0.531622  &  0.396144  &  0.767606  \\
0.417808  &  0.550841  &  0.812506  \\
0.133049  &  0.582132  &  0.812506  \\
0.449110  &  0.866873  &  0.812505  \\
\end{tabular}

\begin{tabular}{lll}
0.809747  &  0.289576  &  0.826178  \\
0.479823  &  0.190171  &  0.826182  \\
0.710357  &  0.520106  &  0.826178  \\
0.152198  &  0.175704  &  0.828662  \\
0.023504  &  0.847791  &  0.828659  \\
0.824271  &  0.976463  &  0.828662  \\
0.175148  &  0.467336  &  0.877220  \\
0.292165  &  0.824822  &  0.877218  \\
0.532660  &  0.707820  &  0.877223  \\
0.178921  &  0.146062  &  0.892626  \\
0.853928  &  0.032852  &  0.892623  \\
0.967142  &  0.821074  &  0.892623  \\
0.825250  &  0.323772  &  0.901161  \\
0.498494  &  0.174711  &  0.901163  \\
0.676213  &  0.501493  &  0.901164  \\
0.360515  &  0.519328  &  0.934324  \\
0.158751  &  0.639433  &  0.934322  \\
0.480632  &  0.841212  &  0.934320  \\
0.821076  &  0.316239  &  0.965366  \\
0.495135  &  0.178916  &  0.965368  \\
0.683754  &  0.504851  &  0.965369  \\
0.175718  &  0.151063  &  0.970353  \\
0.975282  &  0.824281  &  0.970351  \\
0.848927  &  0.024688  &  0.970350  \\
0.169691  &  0.468972  &  0.991280  \\
0.530983  &  0.700771  &  0.991273  \\
0.299195  &  0.830259  &  0.991271  \\
0.154371  &  0.173118  &  0.463715  \\
0.826889  &  0.981246  &  0.463716  \\
0.018765  &  0.845643  &  0.463718  \\
0.479894  &  0.190305  &  0.466914  \\
0.809699  &  0.289583  &  0.466915  \\
0.710429  &  0.520116  &  0.466914  \\
0.134153  &  0.594419  &  0.480908  \\
0.405593  &  0.539708  &  0.480915  \\
0.460316  &  0.865861  &  0.480912  \\
0.520707  &  0.719945  &  0.182196  \\
0.199219  &  0.479313  &  0.182199  \\
0.280076  &  0.800781  &  0.182199  \\
0.858664  &  0.391004  &  0.188804  \\
0.532351  &  0.141360  &  0.188801  \\
0.609007  &  0.467673  &  0.188802  \\
0.408728  &  0.539380  &  0.239805  \\
0.130680  &  0.591297  &  0.239803  \\
0.460624  &  0.869335  &  0.239799  \\
0.799239  &  0.281610  &  0.250515  \\
0.482376  &  0.200785  &  0.250511  \\
0.718424  &  0.517669  &  0.250513  \\
\end{tabular}

\begin{tabular}{lll}
0.158471  &  0.168703  &  0.258094  \\
0.010251  &  0.841564  &  0.258092  \\
0.831341  &  0.989799  &  0.258091  \\
0.173865  &  0.466966  &  0.303449  \\
0.293153  &  0.826173  &  0.303452  \\
0.533049  &  0.706871  &  0.303450  \\
0.177099  &  0.149147  &  0.322073  \\
0.850872  &  0.027942  &  0.322072  \\
0.972076  &  0.822925  &  0.322072  \\
0.822699  &  0.318974  &  0.327476  \\
0.496294  &  0.177308  &  0.327479  \\
0.681047  &  0.503704  &  0.327477  \\
0.363312  &  0.520878  &  0.360276  \\
0.157627  &  0.636721  &  0.360277  \\
0.479154  &  0.842387  &  0.360277  \\
0.823362  &  0.320333  &  0.391727  \\
0.497034  &  0.176674  &  0.391727  \\
0.679706  &  0.503000  &  0.391728  \\
0.177735  &  0.147436  &  0.399660  \\
0.969741  &  0.822272  &  0.399663  \\
0.852567  &  0.030264  &  0.399663  \\
0.174022  &  0.467053  &  0.417093  \\
0.532929  &  0.706917  &  0.417097  \\
0.293094  &  0.825998  &  0.417097  \\
0.162178  &  0.163059  &  0.034470  \\
0.836963  &  0.999212  &  0.034470  \\
0.000812  &  0.837827  &  0.034471  \\
0.485097  &  0.193684  &  0.041797  \\
0.708521  &  0.514932  &  0.041794  \\
0.806332  &  0.291456  &  0.041798  \\
0.378404  &  0.515394  &  0.054657  \\
0.136917  &  0.621616  &  0.054658  \\
0.484618  &  0.863093  &  0.054657  \\
0.529877  &  0.142596  &  0.103859  \\
0.612667  &  0.470126  &  0.103853  \\
0.857455  &  0.387346  &  0.103854  \\
0.175880  &  0.152053  &  0.113680  \\
0.976165  &  0.824174  &  0.113683  \\
0.847971  &  0.023879  &  0.113678  \\
0.518865  &  0.711841  &  0.111947  \\
0.192918  &  0.481148  &  0.111950  \\
0.288182  &  0.807102  &  0.111950  \\
0.175256  &  0.152683  &  0.177884  \\
0.847344  &  0.022647  &  0.177883  \\
0.977374  &  0.824773  &  0.177884  \\
0.333326  &  0.666676  &  0.216518  \\
0.666663  &  0.333358  &  0.214977  \\
0.000026  &  0.000014  &  0.290324  \\
\end{tabular}

\begin{tabular}{lll}
0.333346  &  0.666662  &  0.322674  \\
0.666688  &  0.333334  &  0.359807  \\
0.333331  &  0.666673  &  0.397915  \\
0.000013  &  0.999991  &  0.431410  \\
0.666656  &  0.333336  &  0.077739  \\
0.333316  &  0.666705  &  0.077495  \\
0.999997  &  0.000043  &  0.145859  \\
0.333324  &  0.666665  &  0.504446  \\
0.999988  &  0.999999  &  0.575295  \\
0.666677  &  0.333348  &  0.609316  \\
0.333339  &  0.666674  &  0.646283  \\
0.666665  &  0.333319  &  0.684615  \\
0.999995  &  0.999986  &  0.716961  \\
0.333312  &  0.666613  &  0.788718  \\
0.666642  &  0.333279  &  0.790867  \\
0.000003  &  0.999986  &  0.861013  \\
0.333338  &  0.666652  &  0.896712  \\
0.666652  &  0.333329  &  0.932967  \\
0.333269  &  0.666674  &  0.972094  \\
0.666694  &  0.333376  &  0.502221  \\
0.000001  &  0.000021  &  0.002024  \\	
\end{tabular}


\newpage
\thispagestyle{empty}
\def\thispagestyle#1{}

\renewcommand\bibpreamble{\vspace{-1.0\baselineskip}} 

\titleformat{\chapter}[hang]{\flushleft\fontseries{b}\fontsize{80}{100}\selectfont}{\fontseries{b}\fontsize{100}{130}\selectfont \textcolor{white}\thechapter\hsp}{0pt}{${}$ \\\Huge\bfseries}[]

\clearpage
\phantomsection
\bibliographystyle{unsrt}
\phantomsection
\addcontentsline{toc}{chapter}{Bibliography}
